\documentclass[12pt]{article} 
\usepackage{amsmath}
\usepackage{amssymb}
\usepackage{authblk}
\usepackage{booktabs}
\usepackage{comment}
\usepackage{mathtools}
\usepackage{acronym}
\acrodef{2C-2D}{two-component-two-dimensional}
\acrodef{DES}{detached eddy simulation}
\acrodef{DNS}{direct numerical simulation}
\acrodef{HPOD}{Hilbert POD}
\acrodef{LES}{large eddy simulation}
\acrodef{MCCD}{multigrid/multipass cross-correlation digital}
\acrodef{PCA}{principal component analysis}
\acrodef{PIV}{particle image velocimetry}
\acrodef{POD}{proper orthogonal decomposition}
\acrodef{SVD}{singular value decomposition}
\acrodef{TKE}{turbulent kinetic energy}
\newcommand{\bs}[1]{\boldsymbol{#1}}
\newcommand{\cov}{\operatorname{cov}}

\newcommand{\mc}[1]{\mathcal{#1}}
\newcommand{\ol}[1]{\overline{#1}}
\newcommand{\sgn}{\operatorname{sgn}}
\newcommand{\var}{\operatorname{var}}
\newcommand{\vps}{\varepsilon}
\newcommand{\wt}[1]{\widetilde{#1}}

\usepackage{xcolor}
\usepackage{vmargin}
\setmarginsrb{15mm}  
             {10mm}  
             {15mm}  
             {10mm}  
             {20pt}  
             { 5mm}  
             {10pt}  
             {10mm}  
\title{Studying propagating turbulent structures in the near wake of a sphere using Hilbert proper orthogonal decomposition}
\author{Shaun Davey, Callum Atkinson and Julio Soria}
\affil{Laboratory for Turbulence Research in Aerospace and Combustion,\\Department of Mechanical and Aerospace Engineering,\\Monash University, Clayton, Victoria 3800, Australia}
\date{}
\begin{document}
\maketitle
\begin{abstract}
  Turbulent flows, despite their apparent randomness, exhibit coherent structures that underpin their dynamics.
  Proper orthogonal decomposition (POD) has been widely used to extract these structures from experimental data.
  Periodic features such as vortex shedding can appear as POD mode pairs in strongly periodic flows, but detecting propagating structures in more complex flows is challenging.
  Hilbert proper orthogonal decomposition (HPOD) addresses this by applying POD to the analytic signal of the turbulent fluctuations, which yields complex modes with a $\pi/2$ phase shift between the real and imaginary components.
  These modes capture propagating structures effectively but introduce spectral leakage from the Hilbert transform used to derive the analytic signal.
  The current work investigates the relationship between the modes of the POD and those of the HPOD on the velocity fluctuations in the wake of a sphere.
  By comparing their outputs, POD mode pairs that correspond to the same propagating structures revealed by HPOD are identified.
  Furthermore, this study explores whether computing the analytic signal of the POD modes can replicate the HPOD modes, offering a more computationally efficient method for determining the pairs of POD modes that represent propagating structures.
  The results show that the pairs of POD modes identified by the HPOD can be determined more efficiently using the Hilbert transform directly on the POD modes.
  This method enhances the interpretive power of POD, enabling more detailed analysis of turbulent dynamics without the need to compute the analytic signal of the entire turbulent fluctuation data.
\end{abstract}
\section{Introduction}\label{sec:Intro}
While turbulence often appears random, many turbulent flows contain underlying structures, such as eddies and vortices.
The scale and nature of these structures vary depending on the flow conditions.
Identifying these structures allows for a more detailed description of the flow, as well as the isolation of the dominant structures within the turbulence based on their contribution to the quantity or phenomenon of interest, such as \ac{TKE} or mixing.
Ranking these structures based on their significance to a given application or study allows the construction of parsimonious reduced-order models, which can be used to reduce the computational intensity of analyses or to reduce noise from experimental data.

The complex flow phenomena present in the flow over bluff bodies, even at low Reynolds numbers, develop with increasing Reynolds number, and remain a subject of ongoing study.
In this respect, the flow over spheres serves as a canonical model for flows over axisymmetric bluff bodies, and is of particular interest for advancing the understanding of these flow phenomena.
The phenomena observed in the wake of a sphere include transition from axisymmetry to planar symmetry, from planar symmetry to asymmetry, the formation and shedding of vortices, the development of shear layer instabilities, and flow separation.
As the flow over a sphere varies significantly with Reynolds number, experimental and numerical investigations have explored a wide range of Reynolds number regimes.

In the context of experimental studies of the flow over spheres, one of the main challenges is fixing the sphere in a wind or water tunnel test section with minimal disturbance of the flow, since the mounting or support structure of the sphere frequently has a significant influence on the flow.
The supports chosen for an experiment should simultaneously provide rigid and stable positioning of the sphere, and minimise the disturbance to the flow, while facilitating probe placement and maintaining unobstructed optical access for measurements.
Some of the supports used in these studies include rigid downstream supports~\cite{achenbach1972experiments}, honeycomb structures with rigid supports upstream of the sphere~\cite{grandemange2014statistical,muyshondt2021experimental}, and wires arranged azimuthally~\cite{sakamoto1990study,jang2008piv}.
The drag on the spheres has been measured using hot-wire probes~\cite{achenbach1972experiments,tyagi2006measurement}, with flow visualisations performed using aluminium dust~\cite{taneda1956experimental}, smoke~\cite{taneda1978visual,jang2008piv}, and dyes~\cite{johnson1999flow,sakamoto1990study,achenbach1974vortex} to provide a qualitative understanding of the recirculation region, flow separation, and vortex shedding in the wake.
Various forms of \ac{PIV}, including planar~\cite{jang2008piv}, stereoscopic~\cite{muyshondt2021experimental}, and tomographic~\cite{david2020flow}, have been used to quantitatively measure the flow around spheres.

While mounting the sphere is not a concern in numerical studies, simulations are subject to distinct challenges.
The problem geometry and the wide range of scales involved require careful consideration of the mesh and solver to accurately resolve the flow at high Reynolds numbers.
Because of this complexity, \ac{LES}~\cite{tomboulides1993direct,tomboulides2000numerical,constantinescu2003les,rodriguez2011direct} and \ac{DES}~\cite{constantinescu2003les,constantinescu2004numerical} methods have been preferred over the more computationally expensive \ac{DNS}~\cite{tomboulides1993direct,rodriguez2011direct,rodriguez2013flow,rodriguez2019fluid} for resolving the primary features of the flow at high Reynolds numbers.
In conjunction with the advancement of computational capabilities, these models have expanded the scope of numerical investigations of the flow over spheres to large Reynolds numbers.

Experimental and numerical studies have examined the development of the flow over spheres over a wide range of Reynolds number regimes.
At low Reynolds number, $Re_D < 20$, the flow in the wake is laminar and axisymmetric, and recirculation is absent~\cite{taneda1956experimental,tomboulides2000numerical}.
A small vortex ring forms near the rear stagnation point around $Re_D \approx 25$~\cite{taneda1956experimental}, and elongates in the streamwise direction with increasing Reynolds number.
This vortex ring begins to oscillate around $Re_D \approx 130$ with a long period, which becomes more prominent at higher Reynolds numbers.
The recirculation region formed behind the sphere, which exhibits reverse flow in the mean flow, extends further downstream of the sphere as Reynolds number increases~\cite{johnson1999flow}.

The axisymmetry in the wake at low Reynolds number is lost due to the formation of a double-threaded wake around $Re_D \approx 210$~\cite{johnson1999flow,tomboulides2000numerical}, which consists of two streamwise vortices of opposite sign that can be seen as dye threads in the wake of a free-falling sphere~\cite{magarvey1965vortices}.
Single-frequency vortex shedding begins around $Re_D \approx 300$~\cite{taneda1956experimental,sakamoto1990study}, and hairpin-shaped vortices form at $Re_D \approx 400$~\cite{achenbach1974vortex}.
The generation of an unsteady random side force, which has been observed in both experimental~\cite{taneda1978visual} and numerical~\cite{yun2006vortical} studies, leads to the loss of planar symmetry in the wake within the range of $400 < Re_D < 500$~\cite{taneda1956experimental}.

For $Re_D \gtrapprox 800$, the small-scale Kelvin-Helmholtz instability in the shear layer at the edge of the recirculation region appears as axisymmetric vortex shedding~\cite{constantinescu2004numerical,chomaz1993structure}.
This is the higher-frequency of two instability modes identified in the wake of the sphere by Kim \& Durbin~\cite{kim1988observations}, with the lower-frequency mode arising from large-scale vortex shedding.
Small-scale vortical structures are shed from the main vortical structure around $Re_D \approx 1000$~\cite{yun2006vortical}, and the recirculation region behind the sphere begins to shrink with increasing Reynolds number.
Periodic fluctuations are present in the wake for significantly higher Reynolds numbers, $Re_D = O(10^6)$~\cite{achenbach1974vortex}.

Since its first application to turbulent flows by Lumley~\cite{lumley1967structure}, \ac{POD} has been widely used for identifying and analysing structures in turbulence~\cite{sirovich1987turbulence,berkooz1993proper}.
As \ac{POD} does not require temporal resolution of the flow and can thus be applied to non-time-resolved data, it is a robust yet straightforward method for identifying the structures within a turbulent flow.
In the case of the flow over a cylinder, the turbulent fluctuations due to the vortices shed from the cylinder can be identified in the leading \ac{POD} modes~\cite{oudheusden2005phase,perrin2007phase}.
The energy of these modes is significantly larger than the higher-order modes, and their phase angle is clearly related to that of the pressure fluctuations resulting from the shed vortices~\cite{perrin2007obtaining}.
Similarly, periodic structures in screeching, impinging, and free jets~\cite{edgington2014coherent,weightman2017explanation,weightman2018signatures} are represented by the leading \ac{POD} modes.
Identifying pairs of \ac{POD} modes which represent periodic structures enables the \textit{a posteriori} phase-averaging of experimental data collected asynchronously or at a low sampling frequency relative to these structures.
However, the complexity of the flow over a sphere in this regime makes identification of periodic structures in the turbulent fluctuations challenging, as \ac{POD} modes representing different phases of a structure are not inherently linked by the decomposition.

The analytic signal of the turbulent fluctuations, which is defined using the Hilbert transform, can be decomposed into complex modes that include phase information using \ac{POD}~\cite{horel1984complex}.
While the Hilbert transform is performed in time for time-resolved data, it can also be performed in an analogous spatial direction when the data is not time-resolved~\cite{raiola2024advecting,raiola2024jet,kriegseis2021hilbert}.
This results in modes emphasising structures which are periodic in the chosen direction, with the streamwise direction being the most appropriate choice to identify structures which propagate downstream from the sphere.
Although a useful method for identifying propagating structures, \ac{HPOD} introduces spectral leakage to the modes due to the Hilbert transform being derived from the Fourier transform, which inherently assumes periodicity.

Raiola \& Kriegseis~\cite{raiola2025hilbert} demonstrated the ability of spatial \ac{HPOD} to extract spatiotemporally coherent wavepackets from flows with significant advecting structures.
Their application of \ac{HPOD} to DNS of a laminar vortex street in the wake of a cylinder at $Re_D = 100$ showed that the spatial \ac{HPOD} provides a more compact decomposition than \ac{POD}.
In this case, the first two POD modes representing the vortex street were captured by a single \ac{HPOD} mode.
In this example, the leading mode of the spatial \ac{HPOD} was able to capture the first two \ac{POD} modes even when the data were shuffled in time and the dataset reduced in size, whereas the modes of the conventional time-based \ac{HPOD} converged towards the \ac{POD} modes, and no longer captured paired \ac{POD} modes in a single \ac{HPOD} mode.
They further applied \ac{HPOD} to \ac{LES} of a turbulent jet at $Re_D = 10^6$~\cite{towne2023database}, showing that the spatial \ac{HPOD} could still identify advecting wavepackets in a flow with a broader range of scales.
Finally, \ac{HPOD} was applied to \ac{PIV} data from a turbulent subsonic jet~\cite{raiola2019dynamic} at $Re = 33,000$, which lacked the temporal resolution afforded by the \ac{LES} data. Despite this limitation, the spatial \ac{HPOD} was able to extract spatial modes similar to those identified in the \ac{LES} data.

The current work applies both \ac{POD} and \ac{HPOD}, with the Hilbert transform performed in the streamwise direction following the methodology of~\cite{raiola2025hilbert}, to experimentally measured velocity fluctuations in the wake of a sphere at $Re_D = 7780$.
Modes resulting from each decomposition are compared, and correlations between the \ac{POD} and \ac{HPOD} modes are used to identify \ac{POD} mode pairs representing propagating structures.
A novel approach is introduced in which the Hilbert transform is applied directly to the \ac{POD} modes, allowing mode pairs associated with periodic or propagating structures to be identified without performing \ac{POD} on the complex analytic signal of the turbulent fluctuations.
This method significantly reduces the computational cost of identifying propagating structures and preserves the energy ranking of the modes relative to the original data.
The identified \ac{POD} mode pairs are then used to phase-average the velocity fluctuations in the wake of the sphere to examine the propagating structures, while phase-averaging based on \ac{HPOD} modes is also performed to demonstrate the influence of \ac{HPOD} on the identified structures due to its emphasis on propagating structures and inherent assumption of periodicity.

Section~\ref{sec:Theory} outlines the theoretical aspects of \ac{POD} and \ac{HPOD} and the rationale for applying the Hilbert transform directly to the \ac{POD} modes.
Section~\ref{sec:Method} provides details of the experimental facility and measurement methodology.
Section~\ref{sec:Results} presents the analysis and results of the \ac{POD} and \ac{HPOD}, the pairing of \ac{POD} modes using \ac{HPOD} modes, the application of the Hilbert transform directly to the \ac{POD} modes, and the resulting phase averages of the turbulent fluctuations.
Section~\ref{sec:Conclusions} summarises the work, highlights the key findings, and provides closing remarks.

\section{Theoretical Considerations}\label{sec:Theory}
\subsection{Proper Orthogonal Decomposition}
\ac{POD} decomposes statistically independent snapshots of the random field $\bs{f}(\bs{x},t)$ into orthonormal spatial modes $\psi_k(\bs{x})$ with temporal coefficients $a_k(t)$, which are ranked by their associated energy $\lambda_k$ in descending order, such that
\begin{equation}\label{eq:fxt}
  \bs{f}(\bs{x},t) = \sum_{k=1}^{K} a_k(t)\psi_k(\bs{x}),
\end{equation}
and is mathematically analogous to \ac{SVD} and \ac{PCA}.
In the application of \ac{POD} to turbulence, $\bs{f}(\bs{x},t)$ represents the turbulent fluctuations, and $\lambda_k$ represents the contribution of the $k^{th}$ mode to the total \ac{TKE} of $\bs{f}(\bs{x},t)$.
A common use for \ac{POD} is the definition of a reduced-order representation of $\bs{f}(\bs{x},t)$ by reconstructing the original data using $\hat{K} < K$ modes
\begin{equation}\label{eq:fxt_hat}
  \hat{\bs{f}}(\bs{x},t) = \sum_{k=1}^{\hat{K}} a_k(t)\psi_k(\bs{x}),
\end{equation}
which represents a portion of the total kinetic energy equal to
\begin{equation}
  TKE = 100 \times \frac{\sum_{k=1}^{\hat{K}}\lambda_k}{\sum_{k=1}^{K}\lambda_k} \ \%. 
\end{equation}

In order to perform \ac{POD}, the two-point correlation of $\bs{f}(\bs{x},t)$ in space
\begin{equation}\label{eq:Cx}
  C(\bs{x},\bs{x}') = \int_{T} \bs{f}(\bs{x},t) \bs{f}(\bs{x}',t) \, dt,
\end{equation}
is calculated, where $T$ denotes the temporal domain.
The spatial modes and the corresponding energy are given by the eigendecomposition of $C(\bs{x},\bs{x}')$
\begin{equation}\label{eq:eigenCx}
  \int_S C(\bs{x},\bs{x}') \psi(\bs{x}') \, d\bs{x}'= \lambda \psi(\bs{x}),
\end{equation}
where $S$ denotes the spatial domain.
The temporal coefficients are computed by projecting $\bs{f}(\bs{x},t)$ onto $\psi(\bs{x})$
\begin{equation}\label{eq:a}
  a(t) = \int_S \bs{f}(\bs{x},t)\psi(\bs{x}) \, d\bs{x}.
\end{equation}

Alternatively, \ac{POD} can be implemented using the method of snapshots, which first computes the temporal coefficients, and subsequently determines the spatial modes.
This approach is more computationally efficient when the number of spatial coordinates in $\bs{f}(\bs{x},t)$ exceeds the number of realisations of the field.
The two-instant temporal correlation
\begin{equation}\label{eq:Ct}
  C(t,t') = \int_{S} \bs{f}(\bs{x},t) \bs{f}(\bs{x},t') \, d\bs{x},
\end{equation}
is used instead of the spatial correlation defined in equation (\ref{eq:Cx}).
The eigendecomposition of $C(t,t')$ 
\begin{equation}\label{eq:eigenCt}
  \int_T C(t,t') a(t') \, dt'= \lambda a(t),
\end{equation}
yields the eigenvalues and corresponding temporal coefficients.
The spatial modes are subsequently obtained by projecting $\bs{f}(\bs{x},t)$ onto $a(t)$
\begin{equation}\label{eq:psi}
  \psi(\bs{x})= \int_T \bs{f}(\bs{x},t)a(t) \, dt.
\end{equation}

When the $i^{th}$ and $j^{th}$ POD modes represent the same periodic structure separated in phase by $\pm \pi/2$, their temporal coefficients can be expressed as functions of the phase angle $\phi_{i,j}$ of the structure as
\begin{equation}
  a_i(t) = \sqrt{2 \lambda_i}\cos(\phi_{i,j}(t)),
\end{equation}
and
\begin{equation}
  a_j(t) = \sqrt{2 \lambda_j}\sin(\phi_{i,j}(t)),
\end{equation}
respectively.
Consequently, the phase angle $\phi_{i,j}(t)$ is given by
\begin{equation}\label{eq:phi_ij}
  \phi_{i,j}(t) = \tan^{-1} \left( \frac{a_j(t)/\sqrt{2 \lambda_j}}{a_i(t)/\sqrt{2 \lambda_i}} \right).
\end{equation}

The phase-averaged representation of the data using the $i^{th}$ and $j^{th}$ \ac{POD} modes at the phase angle $\phi$ is defined as
\begin{equation}\label{eq:PhaseAverageWithPOD}
  \wt{\bs{f}}(\bs{x},\phi) = \sum_{k=1}^K a_k(t) \psi_k(\bs{x}) \, | \, \phi_{i,j}(t)=\phi
\end{equation}
where $\phi_{i,j}(t)$ is defined by equation (\ref{eq:phi_ij}).
Although any two \ac{POD} modes can be employed for phase-averaging $\bs{f}(\bs{x},t)$, not all mode selections yield physically meaningful phase averages.
For strongly dominant periodic fluctuations, the temporal coefficients of the corresponding modes exhibit a circular distribution in the coefficient space~\cite{oudheusden2005phase}.
Large-scale periodic structures can also be identified by comparing the phase angle of the modes to an independent phase measurement of the flow, such as pressure signals~\cite{perrin2007obtaining}.
However, as periodic structures become less prominent, establishing a clear relationship between the corresponding \ac{POD} modes becomes increasingly challenging.

\subsection{Hilbert Proper Orthogonal Decomposition}
\ac{HPOD} incorporates phase information into the modes by replacing $\bs{f}(\bs{x},t)$ with its analytic signal~\cite{horel1984complex}
\begin{equation}\label{eq:f^a}
  \bs{f}^a(\bs{x},t) = \bs{f}(\bs{x},t) + i \mc{H}[\bs{f}(\bs{x},t)],
\end{equation}
where $\mc{H}$ denotes the Hilbert transform of $\bs{f}(\bs{x},t)$ with respect to time
\begin{equation}\label{eq:H}
  \mc{H}\big[ \bs{f}(\bs{x},t) \big] = \frac{1}{\pi} \int_{-\infty}^{\infty} \frac{\bs{f}(\bs{x},\tau)}{t-\tau} \, d\tau.
\end{equation}
The Hilbert transform is equivalent to a $\mp \pi/2$ phase shift of the fundamental frequency components of $\bs{f}(\bs{x},t)$ in Fourier space~\cite{thomas1969introduction}
\begin{equation}\label{eq:HilbertInFourier}
  \mc{H}\big[ \bs{f}(\bs{x},t) \big] = \mc{F}^{-1}\Big[-i \sgn(f) \mc{F}\big[ \bs{f}(\bs{x},t) \big]\Big],
\end{equation}
where $\mc{F}$ and $\mc{F}^{-1}$ are the Fourier transform and inverse Fourier transform, respectively, and
\begin{equation}
  \sgn(f) =
  \begin{cases*}
              -1 & \text{if } $f < 0$ \\
    \phantom{-}0 & \text{if } $f = 0$ \\
    \phantom{-}1 & \text{if } $f > 0$ \\
  \end{cases*}
\end{equation}
is the sign of the instantaneous frequency $f$ of each term in the Fourier series given by $\mc{F}\big[\bs{f}(\bs{x},t)\big]$.

For time-resolved data, the Hilbert transform is applied with respect to time to compute the analytic signal $\bs{f}^{a}(\bs{x},t)$.
\ac{HPOD} is implemented on time-resolved data by substituting $\bs{f}(\bs{x},t)$ in equations (\ref{eq:Cx}) to (\ref{eq:a}) with its analytic signal.
For flows where coherent structures propagate downstream, the Hilbert transform can alternatively be applied in the streamwise direction
\begin{equation}\label{eq:Hx}
  \mc{H}_{x} \big[ \bs{f}(\bs{x},t) \big] = \frac{1}{\pi} \int_{-\infty}^{\infty} \frac{\bs{f}\big( (\xi,\ldots), t \big)}{x-\xi} \, d\xi.
\end{equation}
This approach uses streamwise propagation as an analogue for temporal evolution, allowing the method to be applied to data that are not time-resolved~\cite{raiola2024advecting,raiola2024jet,kriegseis2021hilbert}.

In this case, the analytic signal of $\bs{f}(\bs{x},t)$ is defined using the streamwise Hilbert transform $\mc{H}_x$
\begin{equation}\label{eq:f^a_x}
  \bs{f}^a_x (\bs{x},t) = \bs{f}(\bs{x},t) + i \mc{H}_{x} \big[ \bs{f}(\bs{x},t) \big],
\end{equation}
and is substituted for $\bs{f}(\bs{x},t)$ in equations (\ref{eq:Ct}) to (\ref{eq:psi}).
The resulting eigenvalues $\lambda^a_k$ represent the contribution of the $k^{th}$ mode to the total \ac{TKE} of the analytic signal and are real-valued.
The spatial modes $\psi^a_k(\bs{x})$ are complex, with the imaginary component of $\psi^a_k(\bs{x})$ corresponding to a $\mp \pi/2$ phase shift of the real component.

The instantaneous amplitude and phase angle of the $k^{th}$ mode are defined by the real $\Re$ and imaginary $\Im$ components of $a^a_k(t)$ as
\begin{equation}\label{eq:|a^a|}
  |a^a_k(t)| = \sqrt{\Re{\big(a^a_k(t)\big)}^2 + \Im{\big(a^a_k(t)\big)}^2 },
\end{equation}
and
\begin{equation}\label{eq:phi^a}
  \phi^a_k(t) = \tan^{-1} \left( \mp\frac{\Im\big(a^a_k(t)\big)} {\Re\big(a^a_k(t)\big)} \right),
\end{equation}
respectively.
The instantaneous amplitude quantifies the strength of the mode at each instant, while the instantaneous phase angle characterises the mode shape, which oscillates between the real and imaginary components of $\psi^a_k(\bs{x})$ at $\phi^a_k(t)=0^\circ$ and $\phi^a_k(t)=\mp \pi/2$, respectively.

The analytic signal can be reconstructed using $\hat{K}$ modes as
\begin{equation}\label{eq:fa_hat}
  \hat{\bs{f}^a} (\bs{x},t) = \sum_{k=1}^{\hat{K}} a^a_k(t) \psi^a_k(\bs{x}),
\end{equation}
where the real part is the reduced-order model of $\bs{f}(\bs{x},t)$ including the modes which contribute most to the \ac{TKE} of the analytic signal.
This reconstruction favours the propagating components of $\bs{f}(\bs{x},t)$, but introduces spectral leakage into the reconstruction due to the assumption of periodicity inherent to Hilbert transform.

\subsection{Hilbert Transform of POD Modes}
Although \ac{POD} modes do not inherently contain temporal information, the streamwise Hilbert transform can be applied directly to each mode.
Thus, the analytic signal of the \ac{POD} mode $\psi_k(\bs{x})$ is defined as
\begin{equation}
  {(\psi_k)}^a(\bs{x}) = \psi_k(\bs{x}) + i\mc{H}_x \big[\psi_k(\bs{x})\big].
\end{equation}
Substituting the analytic signal of the modes for the \ac{POD} modes in equation (\ref{eq:fxt}) yields
\begin{equation}
  \sum_{k=1}^K a_k(t) \Big( {(\psi_k)}^a(\bs{x}) \Big) = \sum_{k=1}^K a_k(t) \Big( \psi_k(\bs{x}) + i\mc{H}_x \big[ \psi_k(\bs{x}) \big] \Big),
\end{equation}
which can be expanded to
\begin{equation}
  \sum_{k=1}^K a_k(t) \Big( {(\psi_k)}^a(\bs{x}) \Big) = \sum_{k=1}^K a_k(t) \psi_k(\bs{x}) + \sum_{k=1}^K a_k(t) i \mc{H}_x \big[ \psi_k(\bs{x}) \big].
\end{equation}

Since the Hilbert transform is a linear operator, the expression can be rewritten as
\begin{equation}
  \begin{aligned}
    \sum_{k=1}^K a_k(t) \Big( {(\psi_k)}^a(\bs{x}) \Big)
    & = \sum_{k=1}^K a_k(t)\psi_k(\bs{x}) + i \mc{H}_x \Big[ \sum_{k=1}^K  a_k(t)\psi_k(\bs{x}) \Big] \\
    & = \bs{f}(\bs{x},t) + i\mc{H}_x[\bs{f}(\bs{x},t)]                                                \\
    & = \bs{f}^a_x(\bs{x},t),                                                                         \\
  \end{aligned}
\end{equation}
which is the analytic signal of $\bs{f}(\bs{x},t)$.

Therefore, applying the Hilbert transform to each \ac{POD} mode of $\bs{f}(\bs{x},t)$ individually and substituting the resulting analytic signals for the original modes enables the reconstruction of $\bs{f}^a(\bs{x},t)$.
However, the analytic signals of the \ac{POD} modes are not equal to the \ac{HPOD} modes, as the decomposition of the analytic signal of $\bs{f}(\bs{x},t)$ ranks \ac{HPOD} modes by their energy contribution to $\bs{f}^a(\bs{x},t)$ rather than to $\bs{f}(\bs{x},t)$.
Consequently, leading POD modes may contain non-propagating features that are relegated to higher-order \ac{HPOD} modes.
While this approach is not equivalent to \ac{HPOD}, the tendency for propagating structures to be decomposed into paired POD modes---or real and imaginary parts of \ac{HPOD} modes~\cite{raiola2025hilbert}---suggests that the Hilbert transform of a POD mode representing a propagating structure should closely match the \ac{POD} mode corresponding to the $\pi/2$ phase-shift of the same structure.

\section{Methodology}\label{sec:Method}
\subsection{Experimental Facility}
\begin{figure}
  \centering
  \includegraphics[width=5.3in]{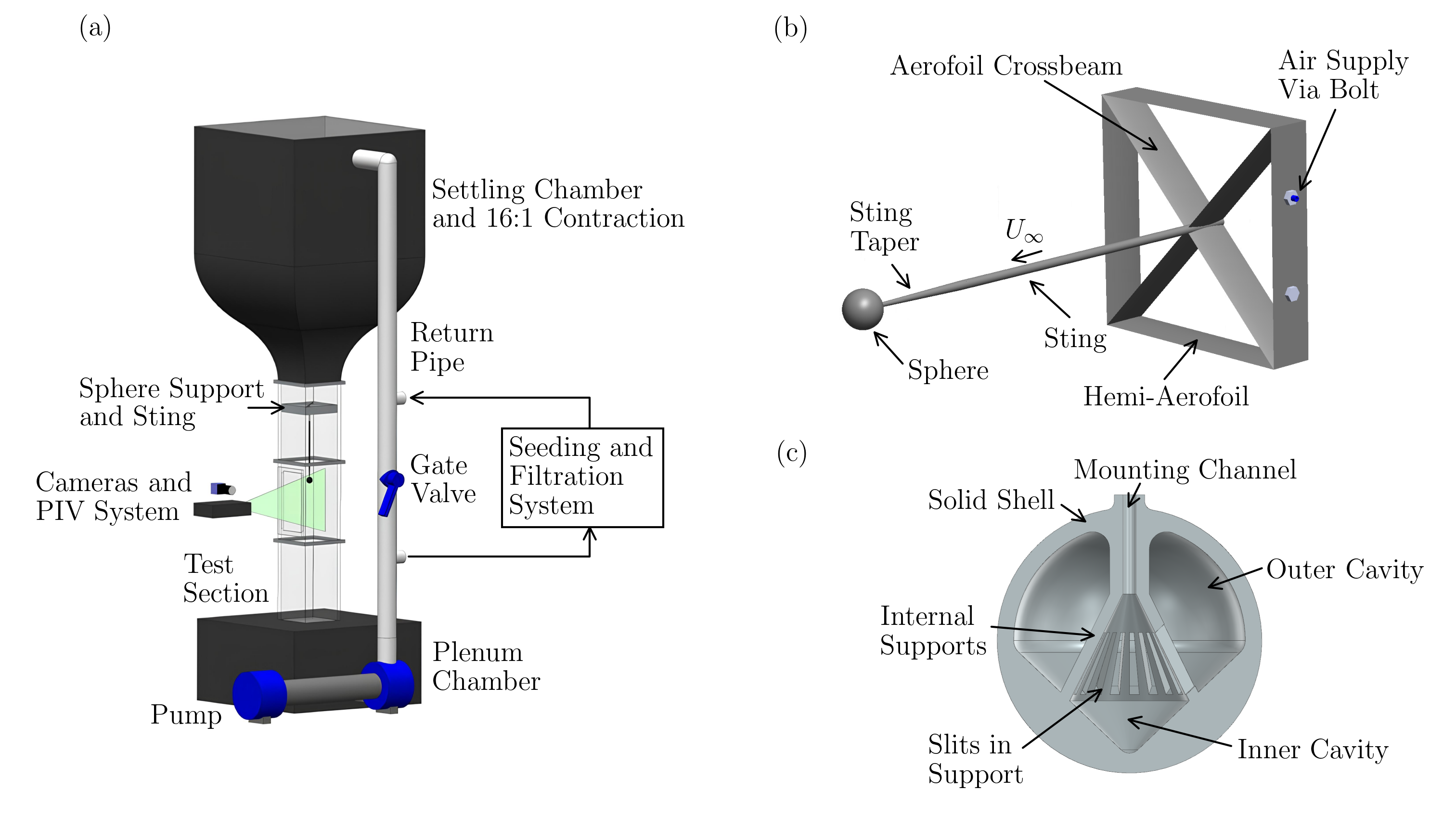}
  \caption{(a) Vertical water tunnel facility, (b) close-up of the mounting structure and sphere, and (c) 3-D printed sphere design~\cite{davey2025experimental}.}\label{fig:VWT}
\end{figure}

Experiments were performed in a vertical water tunnel, the test section of which is 1.5 m long with a cross-section of 0.25 m $\times$ 0.25 m, shown in figure~\ref{fig:VWT}a~\cite{buchner2015measurements,buchner2017stability,davey2025experimental}.
The settling chamber is connected to the top of the test section by a 16:1 contraction, which incorporates a honeycomb and screens to reduce the turbulence intensity and scale in the tunnel test section.
Water is pumped from the plenum chamber below the test section up to the settling chamber such that the flow through the test section is downwards.
The walls of the test section are made of 15 mm thick clear acrylic.
A removable panel is located on one wall for access to the test section, and is set flush to the internal side of the wall to minimise asymmetry in the tunnel.
Seeding and filtration of the tunnel is performed via an auxiliary circuit, which is isolated during experiments.
A sting of diameter 9.3 mm is mounted to a crossbeam of thin aerofoil cross-sections above the test section.
The sting tapers over the last 100 mm to a diameter of 4.65 mm and terminates in a threaded rod used to mount the sphere.
A close-up view of the sphere mounting structure is shown in figure~\ref{fig:VWT}b.

A sphere with a diameter of $D=40$ mm was fabricated using a Phrozen Sonic Mighty 8k 3D printer, featuring a layer height of 35 \textmu\@m and a planar resolution of 22 \textmu\@m $\times$ 22 \textmu\@m, as shown in figure~\ref{fig:VWT}c.
The sphere has a shell thickness of 2.5 mm and an external boss at the top for mounting to the threaded rod on the end of the sting.
The boss begins parallel to the sting and smoothly transitions to the sphere surface.
The sphere is supported internally by a conical structure, which features slits to prevent uncured resin from being trapped in the cavities during printing.

\subsection{Experimental Method}
Flow measurements in the wake of the sphere were conducted at a freestream velocity of $U_\infty=200$ mm/s, corresponding to a Reynolds number of $Re_D=7780$.
Under these conditions, the background turbulence is 2\%, and the mean convective time of the flow
\begin{equation}\label{eq:t_c}
  t_c = \frac{D}{U_\infty},
\end{equation}
was 0.2 seconds~\cite{davey2025experimental}.
The flow was seeded with 11 \textmu\@m glass spheres, and single-exposed particle image pairs were recorded using a PCO Panda 26 DS camera with a 5120 $\times$ 5120 px array of 2.5 \textmu\@m $\times$ 2.5 \textmu\@m pixels.
A Zeiss lens with a fixed 50 mm focal length was used to capture a 190 mm $\times$ 190 mm field of view, resulting in a spatial resolution of 37 \textmu\@m/px.
The field of view began just behind the trailing edge of the sphere, to avoid reflections off the surface, and extended to 5.2 diameters downstream of the sphere.
The field of view was aligned with the centreline of the sphere, and extended to 2.3 diameters either side of the centreline in the transverse direction.
The f-stop was set to 2.8, yielding a depth of field of 5.0 mm.
The diffraction-limited minimum image diameter and expected particle image diameter for an 11 \textmu\@m particle at these settings were 3.9 \textmu\@m and 1.6 px, respectively.

Illumination was provided by 7 ns pulses from a pair of 400 mJ Spectra Physics Nd:YAG lasers.
The beams were aligned collinearly and shaped into a sheet aligned to the tunnel centreline and the corresponding meridian of the sphere.
The thickness of the laser sheet throughout the field of view was approximately 300 \textmu\@m.
Image pairs were captured with a separation of $\Delta t = 1$ ms at an acquisition rate of 1 Hz ($\Delta T = 1$ s), using a BeagleBone Black pulse generator~\cite{fedrizzi2015application} to synchronise the camera and laser pulses.

\subsection{Data Processing}
Mean background images of the first and second exposures were subtracted from the corresponding particle images.
The instantaneous streamwise and transverse velocities, $u(x,y,t)$ and $v(x,y,t)$, were determined from the particle image pairs using \ac{2C-2D} \ac{MCCD} \ac{PIV}~\cite{willert1991digital,soria1996investigation}.
The final interrogation window size was 64 px $\times$ 32 px, with a grid spacing of 32 px $\times$ 16 px in the streamwise $x$ and transverse $y$ directions, respectively.
This aspect ratio was selected to capture the higher streamwise velocity component and the higher velocity gradients in the transverse direction, while ensuring each interrogation window contained approximately 10 particles common to both images of each pair.
The quality of the cross-correlation was increased through the application of an 8-point Hart filter.
A maximum displacement limit of 20 px and normalised local median filter~\cite{westerweel2005universal}, with a threshold of 2 standard deviations, were applied to the raw displacement vectors.
The validation criteria resulted in a rejection of 2.3\% of vectors, which were replaced by an interpolation using a second-order polynomial fit of their 13 nearest neighbours~\cite{soria1996investigation}.
A total of $N_T = 12,600$ two-component velocity vector fields, each containing 50,000+ vectors, were determined.
The instantaneous velocity fluctuations $u'(x,y,t)$ and $v'(x,y,t)$ were calculated by subtracting the mean velocity field, $\ol{u}(x,y)$ and $\ol{v}(x,y)$, from the instantaneous velocity field, $u(x,y,t)$ and $v(x,y,t)$.

The 95\% confidence interval of the \ac{2C-2D} \ac{MCCD}\ac{PIV} algorithm used is 0.06 pixels~\cite{soria1996investigation}, corresponding to a standard error of $\sigma_{\vps_u}$ = 0.03 pixels.
This translates to a standard error of $1.1 \times 10^{-3}$ m/s in the instantaneous velocities.
The uncertainty of the mean velocity arising from the random nature of the turbulent fluctuations can be estimated by including the mean streamwise Reynolds stress~\cite{benedict1996towards,sun20252c2d}
\begin{equation}\label{eq:unc_U}
  \sigma_{\ol{u}} = \sqrt{\frac{\ol{u'u'} + \sigma^2_{\vps_u}}{N_T}},
\end{equation}
where $\sigma_{\vps_{u}}$ is the standard error of the instantaneous streamwise velocity measurements.
As the maxima of both $\ol{u'u'}$ and $\ol{v'v'}$ are of the order of $0.05U_\infty^2$~\cite{davey2025experimental}, a conservative estimate of the uncertainty in the mean velocity components is $4.0 \times 10^{-4}$ m/s.
The uncertainty in the fluctuations is thus
\begin{equation}\label{eq:unc_uf}
  \sigma_{u'} = \sqrt{\sigma^2_{\ol{u}} + \sigma^2_{\vps_u}},
\end{equation}
which is equal to $1.2 \times 10^{-3}$ m/s.
The key experimental and processing parameters are summarised in table~\ref{tab:ExperimentalParameters}.

\begin{table}
  \centering
  \begin{tabular}{cccc}
    \toprule
    \bf{Parameter}                                 & \bf{Symbol}     & \bf{Units}                & \bf{Relative Units}     \\
    \midrule
    Freestream velocity                            & $U_\infty$      & 200 mm/s                  & $-$                     \\
    Sphere diameter                                & $D$             & 40 mm                     & $-$                     \\
    Reynolds number                                & $Re_D$          & 7780                      & $-$                     \\
    Convective time                                & $t_c$           & 0.2 s                     & $-$                     \\
    Time between laser pulses                      & $\Delta t$      & 1 ms                      & $0.005t_c$              \\
    Time between velocity fields                   & $\Delta T$      & 1 s                       & $5t_c$                  \\
    Depth of field                                 & $-$             & 5.0 mm                    & $0.125D$                \\
    Laser sheet thickness                          & $-$             & 0.3 mm                    & $0.0075D$               \\
    Diffraction-limited particle diameter          & $-$             & 3.9 \textmu\@m            &                         \\
    Spatial resolution                             & $SR$            & 37 \textmu\@m             & $-$                     \\
    Grid spacing                                   & $-$             & 32 px $\times$ 16 px      & $0.03D \times 0.015D$   \\
    Interrogation window                           & $-$             & 64 px $\times$ 32 px      & $0.06D \times 0.06D$    \\
    Field of view                                  & $-$             & 190 mm $\times$ 190 mm    & $4.7D \times 4.7D$      \\
    Number of 2C-2D velocity fields                & $N_T$           & 12,600                    & $-$                     \\
    Velocity vectors per field                     & $N_S$           & $\phantom{-}$50,000+      & $-$                     \\
    Velocity vector validation rate                & $-$             & 97.7\%                    &                         \\
    Instantaneous velocity fluctuation uncertainty & $\sigma_{u'}$   & $1.2 \times 10^{-3}$ m/s  & $0.006U_\infty$         \\
    \bottomrule
  \end{tabular}
  \caption{Summary of experimental and processing parameters.}\label{tab:ExperimentalParameters}
\end{table}

\section{Analysis and Results}\label{sec:Results}
\subsection{POD and HPOD Modes}
\ac{POD} was implemented by arranging the turbulent fluctuations into the matrix
\begin{equation}\label{eq:Xmn}
  \bs{X} = \begin{bmatrix}
    u'(\bs{x}_1,t_1)     & \cdots           & u'(\bs{x}_{N_S},t_1)     & v'(\bs{x}_1,t_1)     & \cdots           & v'(\bs{x}_{N_S},t_1)     \\
    \vdots               & u'(\bs{x}_n,t_m) & \vdots                   & \vdots               & v'(\bs{x}_n,t_m) & \vdots                   \\
    u'(\bs{x}_1,t_{N_T}) & \cdots           & u'(\bs{x}_{N_S},t_{N_T}) & v'(\bs{x}_1,t_{N_T}) & \cdots           & v'(\bs{x}_{N_S},t_{N_T}) \\
  \end{bmatrix}
\end{equation}
where $N_S$ denotes the number of velocity vectors per velocity field and $\bs{x}_n={(x,y)}_n$ represents the spatial coordinates of the $n^{th}$ vector, with fluctuating components $u'$ and $v'$ corresponding to the streamwise and transverse directions, respectively.
Since $N_S > N_T$, the method of snapshots was employed to perform the \ac{POD}.
The two-instant correlation of $\bs{X}$ is expressed in matrix form as
\begin{equation}\label{eq:CT_matrix}
  \bs{C} = \bs{X} \bs{X}^T,
\end{equation}
the eigendecomposition of which
\begin{equation}\label{eq:eigCT_matrix}
  \bs{C}\bs{\Phi} = \lambda \bs{\Phi},
\end{equation}
yields the eigenvectors $\bs{\Phi}$ and eigenvalues $\lambda$.
The temporal coefficients are obtained from $\bs{\Phi}$
\begin{equation}\label{eq:a_matrix}
  a_k(t_m) = \bs{\Phi}_{m,k},
\end{equation}
and the spatial modes are given by projecting $\bs{X}$ onto $\bs{\Phi}$
%
\begin{equation}\label{eq:psi_matrix}
  \psi_k(\bs{x}_n) = {(\bs{\Phi}^T \bs{X})}_{k,n}.
\end{equation}

As the Fourier transform---and thus the Hilbert transform---assumes periodic signals, the matrix $\bs{X}$ was weighted in the streamwise direction using a Hann window~\cite{blackman1958measurement} prior to performing \ac{HPOD} to reduce spectral leakage in the modes.
The weight at each streamwise index is defined as
\begin{equation}\label{eq:HannWindow}
  w_\textrm{Hann}(x) = \tanh\!\left( \alpha \sin^2( \pi n/N ) \right),
\end{equation}
where $\alpha$ is a scaling parameter that controls the steepness of the weighting taper towards the domain edges, $n$ is the index of the streamwise location and $N$ is the total number of streamwise locations.
A value of $\alpha=10$ was used to provide a relatively steep taper to zero at the domain edges while minimising the effect on the interior.
The weighted analytic signal of $\bs{X}$
\begin{equation}\label{eq:HannHPOD}
  \bs{X}^a = w_\textrm{Hann}(x) \bs{X} + i\mc{H}_x[w_\textrm{Hann}(x) \bs{X}],
\end{equation}
was determined using equations (\ref{eq:Hx}) and (\ref{eq:f^a_x}), and \ac{HPOD} was performed by substituting $\bs{X}^a$ for $\bs{X}$ in equations (\ref{eq:CT_matrix}) to (\ref{eq:psi_matrix}).
The resulting eigenvalues, temporal coefficients, and spatial modes are denoted by $\lambda^a_k$, $a^a_k(t)$, and $\psi^a_k(\bs{x})$, respectively.

\begin{figure}
  \centering
  \includegraphics[width=5.3in]{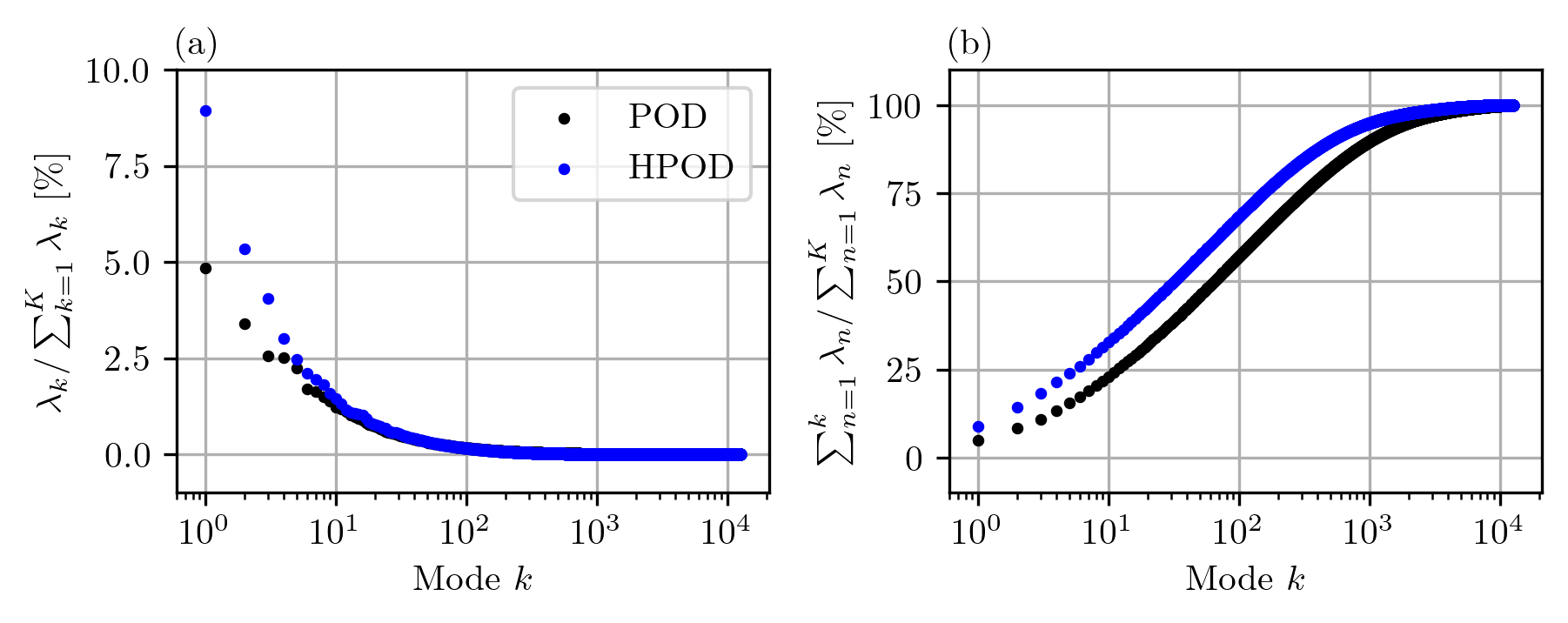}
  \caption{(a) Individual and (b) cumulative \ac{TKE} contribution of the \ac{POD} and \ac{HPOD} modes.}\label{fig:TKE}
\end{figure}

The contribution of the leading \ac{HPOD} modes to the \ac{TKE} of $\bs{X}^a$ is significantly larger than that of the leading \ac{POD} modes to the \ac{TKE} of $\bs{X}$.
The first, second, and third modes of the \ac{HPOD} contribute 8.93\%, 5.36\%, and 4.05\%, respectively, while those of the \ac{POD} contribute 4.85\%, 3.39\%, and 2.55\%, respectively.
However, the difference in the contribution of the higher-order modes to the \ac{TKE} is less pronounced, as shown in figure~\ref{fig:TKE}a.
Consequently, the cumulative contribution to \ac{TKE} of a given number of the \ac{HPOD} modes is more than that of the equivalent number of \ac{POD} modes, as shown in figure~\ref{fig:TKE}b and table~\ref{tab:TKE}.
It should be noted that this is not a one-to-one comparison, as the total \ac{TKE} of the analytic signal $\bs{X}^a$, 453,071 m$^2$/s$^2$, is greater than that of the data $\bs{X}$, 300,542 m$^2$/s$^2$.
This difference arises from the imaginary component of $\bs{X}^a$, which is absent from $\bs{X}$.

\begin{table}
  \centering
  \begin{tabular}{cccc}
    \toprule
    \bf{\ac{TKE} [\%]} & \bf{\ac{POD} Modes} & \bf{\ac{HPOD} Modes} & \bf{Ratio} \\
    \midrule
    25                 &   12                &    6                 & 0.50       \\
    50                 &   65                &   32                 & 0.49       \\
    75                 &  311                &  153                 & 0.49       \\
    90                 & 1027                &  513                 & 0.50       \\
    95                 & 1973                & 1020                 & 0.52       \\
    99                 & 5904                & 3753                 & 0.64       \\
    \bottomrule
  \end{tabular}
  \caption{Number of \ac{POD} modes and \ac{HPOD} modes required to meet 25\% to 99\% of the \ac{TKE} of the data $\bs{X}$ and its analytic signal $\bs{X}^a$, respectively, and the ratio of \ac{HPOD} modes to \ac{POD} modes required to meet each proportion of \ac{TKE}.}\label{tab:TKE}
\end{table}

\begin{figure}
  \centering
  \includegraphics[width=5.3in]{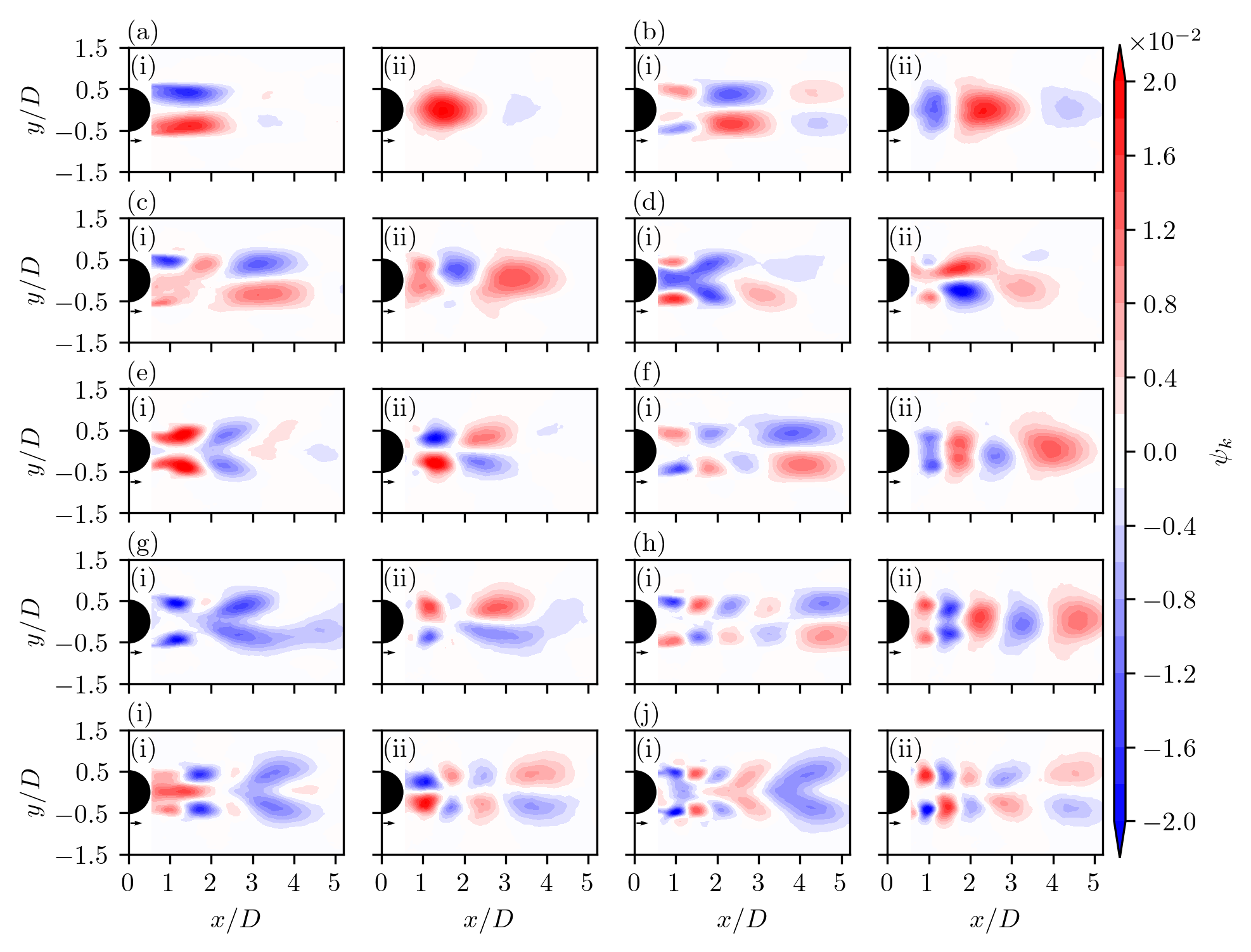}
  \caption{(i) Streamwise and (ii) transverse components of the first ten \ac{POD} modes.}\label{fig:POD_Modes}
\end{figure}

The first \ac{POD} mode consists of strong fluctuations directly behind the sphere at $x/D < 2.5$, with substantially weaker fluctuations present around $x/D = 3.5$.
The streamwise fluctuations, shown in figure~\ref{fig:POD_Modes}a\@(i), are opposite in sign across the centreline, where they vanish, while the transverse fluctuations, shown in figure~\ref{fig:POD_Modes}a\@(ii), alternate in sign along the streamwise direction and are maximal at the centreline.
The second \ac{POD} mode, the streamwise and transverse components of which are shown in figures~\ref{fig:POD_Modes}b\@(i) and~\ref{fig:POD_Modes}b\@(ii), respectively, appears similar to the first mode, with the structures shifted downstream.
The structures present at $x/D < 2.5$ in the first mode are now located at $1.5 < x/D < 3.5$, and new structures with opposite sign are present at $x/D < 1.5$, suggesting that these modes represent a propagating structure of fluctuations with opposing sign.
Although the third \ac{POD} mode, shown in figure~\ref{fig:POD_Modes}c, is asymmetric in the measured plane, it is similar in structure to the first and second modes, suggesting that the structure is not always planar-symmetric.
The fourth and fifth \ac{POD} modes, shown in figures~\ref{fig:POD_Modes}d and~\ref{fig:POD_Modes}e, respectively, both exhibit symmetric V-shaped structures in the streamwise fluctuations and antisymmetric structures in the transverse fluctuations.
Although the streamwise and transverse components of the fifth \ac{POD} mode appear more symmetric and antisymmetric, respectively, than those of the fourth mode, it still resembles the fourth mode shifted downstream.

The sixth and eighth \ac{POD} modes, shown in figures~\ref{fig:POD_Modes}f and~\ref{fig:POD_Modes}h, respectively, are similar in structure to the first and second \ac{POD} modes but with a shorter wavelength.
The seventh \ac{POD} mode superficially resembles the fourth mode in its transverse component, shown in figure~\ref{fig:POD_Modes}g(ii).
However, the streamwise component, shown in figure~\ref{fig:POD_Modes}g(i), does not exhibit the streamwise periodicity of the other leading modes.
While this mode may represent an asymmetric aspect of the structure represented by the fourth and fifth modes, it is less clearly related to these modes than the third mode appears to be to the first and second modes.
The ninth and tenth \ac{POD} modes exhibit V-shaped structures in their streamwise components, shown in figures~\ref{fig:POD_Modes}i(i) and~\ref{fig:POD_Modes}j(i), respectively, similar to the fourth and fifth modes for $x/D > 2.5$, which increase in wavelength further downstream.
These modes differ in shape from the lower-order modes upstream of $x/D = 2$, where the structures have an aspect ratio closer to unity than those in the higher-order modes.
The transverse components of the ninth and tenth modes, shown in figures~\ref{fig:POD_Modes}i(ii) and~\ref{fig:POD_Modes}j(ii), respectively, resemble those of the fourth and fifth modes with a short wavelength.
The wavelength of these modes also increases in the downstream direction.

\begin{figure}
  \centering
  \includegraphics[width=5.3in]{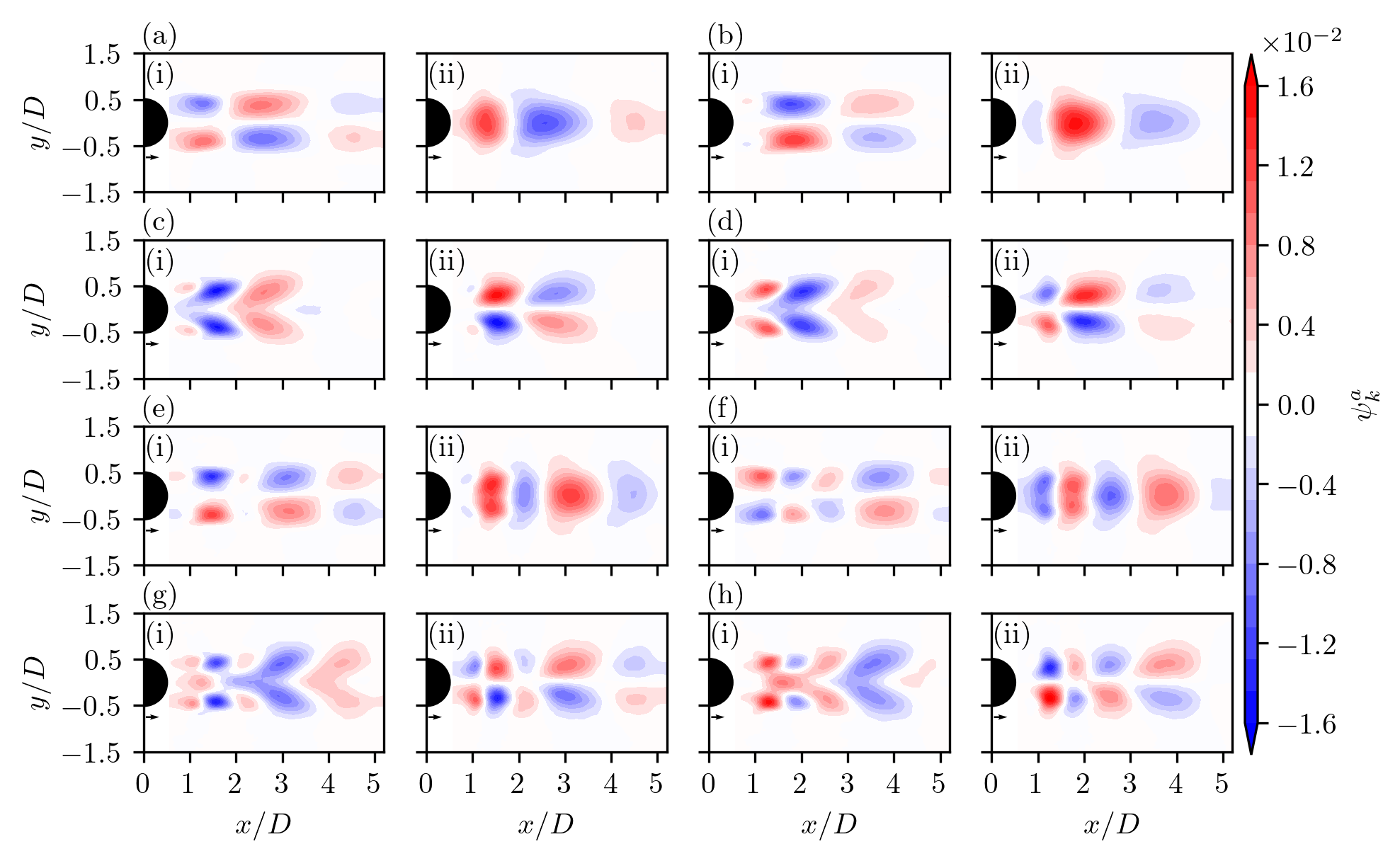}
  \caption{(i) Streamwise and (ii) transverse components of the first four \ac{HPOD} modes.
  Real parts are shown in (a), (c), (e), and (g) and corresponding imaginary parts are shown in (b), (d), (f), and (h)}\label{fig:HPOD_Modes}
\end{figure}

The first \ac{HPOD} mode, the real and imaginary parts of which are shown in figure~\ref{fig:HPOD_Modes}a and~\ref{fig:HPOD_Modes}b, respectively, is qualitatively similar in structure to the first two \ac{POD} modes, shown in figure~\ref{fig:POD_Modes}a and~\ref{fig:POD_Modes}b, respectively.
The streamwise component of the first \ac{HPOD} mode, the real and imaginary components of which are shown in figures~\ref{fig:HPOD_Modes}a(i) and~\ref{fig:HPOD_Modes}b(i), respectively, consists of alternating positive and negative regions in the streamwise direction and is antisymmetric across the centreline.
The transverse component, the real and imaginary parts of which are shown in figures~\ref{fig:HPOD_Modes}a(ii) and~\ref{fig:HPOD_Modes}b(ii), respectively, consists of symmetric regions of alternating sign.
This structure is very similar to that of the first two \ac{POD} modes, except for the ends of the measurement domain, where the intensity is reduced by the Hann window.
The small structures near the sphere at $x/D = 1$, in the second \ac{POD} mode, shown in figure~\ref{fig:POD_Modes}b(i), are extended towards the centreline in the real component of the \ac{HPOD} mode, similar to the downstream structures, due to the assumption of periodicity implicit in the Hilbert transform.
In the imaginary part of the \ac{HPOD} mode, shown in figure~\ref{fig:HPOD_Modes}b(i), these structures are less pronounced as their intensity is reduced by the Hann window.
The favouring of propagating structures by \ac{HPOD} also results in an increase in the relative intensity of the downstream structures in figures~\ref{fig:HPOD_Modes}a(i) and~\ref{fig:HPOD_Modes}b(i) compared to those in figures~\ref{fig:POD_Modes}a(i) and~\ref{fig:POD_Modes}b(i).

The streamwise component of the second \ac{HPOD} mode, the real and imaginary parts of which are shown in figures~\ref{fig:HPOD_Modes}c(i) and~\ref{fig:HPOD_Modes}d(i), respectively, features the same V-shaped structure present in the fourth and fifth \ac{POD} modes, shown in figures~\ref{fig:POD_Modes}d(i) and~\ref{fig:POD_Modes}e(i), respectively.
The transverse components of the real and imaginary parts of the second \ac{HPOD} mode are antisymmetric with regions of alternating sign along the streamwise direction, as shown in figures~\ref{fig:HPOD_Modes}c(ii) and~\ref{fig:HPOD_Modes}d(ii), respectively.
While similar to the transverse components of the fourth and fifth \ac{POD} modes, shown in figures~\ref{fig:POD_Modes}d(ii) and~\ref{fig:POD_Modes}e(ii), respectively, the \ac{HPOD} modes are significantly more symmetric and antisymmetric across the centreline than the \ac{POD} modes, due to the accentuation of propagating structures in the \ac{HPOD} modes.

The third \ac{HPOD} mode, the real and imaginary components of which are shown in figures~\ref{fig:HPOD_Modes}e and~\ref{fig:HPOD_Modes}f, respectively, has a similar structure to the sixth and eighth \ac{POD} modes, shown in figures~\ref{fig:POD_Modes}f and~\ref{fig:POD_Modes}h, respectively.
Specifically, it exhibits antisymmetric streamwise velocity regions and symmetric transverse velocity regions of alternating sign along the streamwise direction, with a shorter wavelength than the first \ac{HPOD} mode and the first two \ac{POD} modes.
The fourth \ac{HPOD} mode, the real and imaginary components of which are shown in figures~\ref{fig:HPOD_Modes}g and~\ref{fig:HPOD_Modes}h, has a similar structure to the ninth and tenth \ac{POD} modes, shown in figures~\ref{fig:POD_Modes}i and~\ref{fig:POD_Modes}j, respectively.
The streamwise component of this mode features symmetric ovular regions of alternating sign in the streamwise direction in the near wake, becoming V-shaped structures around $x/D = 2$, while the transverse component features antisymmetric regions of alternating sign in the streamwise direction. 

\subsection{Pairing POD Modes using HPOD Modes}

As $\bs{f}(\bs{x},t)$ is given by the real part of $\bs{f}^a(\bs{x},t)$, it can be expressed using the \ac{HPOD} modes as
\begin{equation}\label{eq:Xfrompsi^a}
  \bs{f}(\bs{x},t) = \Re\big(\bs{f}^a(\bs{x},t) \big) = \sum_{k=1}^K \Re\big( a^a_k(t) \psi^a_k(\bs{x}) \big),
\end{equation}
which can be expanded to
\begin{equation}\label{eq:Xfrompsi^a_expanded}
  \bs{f}(\bs{x},t) = \Re\big(\bs{f}^a(\bs{x},t) \big) = \sum_{k=1}^K \Re\big( a^a_k(t) \big) \Re\big( \psi^a_k(\bs{x}) \big) - \Im\big( a^a_k(t) \big)\Im\big( \psi^a_k(\bs{x}) \big).
\end{equation}
As the real and imaginary parts of $a^a_k(t)$ can be expressed as $|a^a_k(t)|\cos(\phi^a_k(t))$ and $-|a^a_k(t)|\sin(\phi^a_k(t))$, respectively, the shape of the $k^{th}$ \ac{HPOD} mode for the phase angle $\phi^a_k(t)$ is given by
\begin{equation}\label{eq:psi_phi^a_k}
  \frac{\psi^a_k(\bs{x},\phi^a_k(t))}{|a^a_k(t)|} = \cos \big( \phi^a_k(t) \big) \Re\big( \psi^a_k(\bs{x}) \big) + \sin \big( \phi^a_k(t) \big) \Im\big( \psi^a_k(\bs{x}) \big).
\end{equation}
The mode shape of the $k^{th}$ \ac{HPOD} mode matches the $j^{th}$ \ac{POD} mode when
\begin{equation}\label{eq:POD_HPOD_Modes}
  \psi_j(\bs{x}) =
  \begin{bmatrix}
    \cos(\Delta \phi^a_{k,j}) & \sin(\Delta \phi^a_{k,j}) \\
  \end{bmatrix}
  \begin{bmatrix}
    \Re \big( \psi^a_k (\bs{x}) \big)\\
    -\Im \big( \psi^a_k (\bs{x}) \big)\\
  \end{bmatrix},
\end{equation}
where $\Delta\phi_{k,j}^a$ is the phase offset of the $k^{th}$ \ac{HPOD} mode which provides the best match to the $j^{th}$ \ac{POD} mode.
The negative sign on the imaginary part accounts for the negative introduced by the Hilbert transform associated with positive instantaneous frequencies, given in equation (\ref{eq:HilbertInFourier}), which accounts for the phase of the \ac{HPOD} modes being reversed relative to the motion of the structure.

The phase angles at which the $k^{th}$ \ac{HPOD} mode matches each of the \ac{POD} modes can be determined by solving
\begin{equation}\label{eq:AllPOD_HPOD}
  \begin{bmatrix}
    \cos(\Delta\phi^a_{k,1}) & \sin(\Delta\phi^a_{k,1}) \\
    \vdots                   & \vdots                   \\
    \cos(\Delta\phi^a_{k,j}) & \sin(\Delta\phi^a_{k,j}) \\
    \vdots                   & \vdots                   \\
    \cos(\Delta\phi^a_{k,K}) & \sin(\Delta\phi^a_{k,K}) \\
  \end{bmatrix}^T
  =
  {\left(\begin{bmatrix}
    \Re\big(\psi^a_k(\bs{x})\big) \\ -\Im\big(\psi^a_k(\bs{x})\big) \\
  \end{bmatrix}
  \begin{bmatrix}
    \Re\big(\psi^a_k(\bs{x})\big) \\ -\Im\big(\psi^a_k(\bs{x})\big) \\
  \end{bmatrix}^T\right)}^{-1}
  \begin{bmatrix}
    \Re\big(\psi^a_k(\bs{x})\big) \\ -\Im\big(\psi^a_k(\bs{x})\big) \\
  \end{bmatrix}
  \begin{bmatrix}
    \psi_1(\bs{x}) \\ \vdots \\ \psi_j(\bs{x}) \\ \vdots \\ \psi_K(\bs{x}) \\
  \end{bmatrix}^T,
\end{equation}
and
\begin{equation}\label{eq:BestPhi}
  \Delta\phi^a_{k,j} = \tan^{-1} \left( \frac{\sin(\Delta\phi^a_{k,j})}{\cos(\Delta\phi^a_{k,j})} \right).
\end{equation}
The correlation coefficient between the $k^{th}$ \ac{HPOD} mode at $\Delta \phi^a_{k,j}$ and the $j^{th}$ \ac{POD} mode is given by
\begin{equation}\label{eq:Rkj}
  \mc{R}_{k,j}(\Delta\phi^a_{k,j}) = \frac{\cov \big( \psi^a_k(\bs{x},\Delta\phi^a_{k,j}),\psi_j(\bs{x}) \big)}{\sqrt{\var \big( \psi^a_k(\bs{x},\Delta\phi^a_{k,j}) \big) \var \big( \psi_j(\bs{x}) \big)}},
\end{equation}
and the corresponding $R^2_{k,j}$ value is
\begin{equation}\label{eq:R2_kj}
  R^2_{k,j}(\Delta\phi^a_{k,j}) = 1 - \frac{\sum_{\bs{x}} {\Big(\psi^a_k(\bs{x},\Delta \phi_{k,j}^a) - \psi_j(\bs{x}) \Big)}^2}{\sum_{\bs{x}} {\Big( \psi_j(\bs{x}) \Big)}^2},
\end{equation}
where $\bs{x}$ represents the spatial coordinates of the modes.
The $\mc{R}_{k,j}$ and $R_{k,j}^2$ values were calculated for the combined streamwise and transverse components of the modes, as well as individually for each component, denoted by superscripts $^u$ and $^v$ for the streamwise and transverse components, respectively. 
The \ac{HPOD} modes which best match each of the first ten \ac{POD} modes, and the corresponding phase offset $\Delta\phi^a_{k,j}$, correlation coefficient $\mc{R}_{k,j}$, and $R_{k,j}^2$ value, are shown in table~\ref{tab:R_HPOD,POD}.

\begin{table}
  \centering
  \begin{tabular}{c c c c c c c c c}
    \toprule
    \bf{POD Mode}, $j$ & \bf{HPOD Mode}, $k$ & \bf{$\Delta \phi_{k,j}^a$} & \bf{$\mc{R}_{k,j}$} & \bf{$\mc{R}_{k,j}^u$} & \bf{$\mc{R}_{k,j}^v$} & $R^2_{k,j}$ & $(R^2_{k,j})^u$ & $(R^2_{k,j})^v$ \\
    \midrule
    1                  & 1                   & $1.64\pi$                  & 0.88                & 0.86                  & 0.90                  & 0.74        & 0.71            & 0.75            \\
    2                  & 1                   & $1.15\pi$                  & 0.98                & 0.96                  & 0.99                  & 0.95        & 0.92            & 0.96            \\
    3                  & 3                   & $0.65\pi$                  & 0.57                & 0.46                  & 0.66                  & 0.28        & 0.10            & 0.38            \\
    4                  & 2                   & $1.71\pi$                  & 0.86                & 0.85                  & 0.88                  & 0.68        & 0.70            & 0.74            \\
    5                  & 2                   & $1.21\pi$                  & 0.96                & 0.95                  & 0.97                  & 0.84        & 0.87            & 0.83            \\
    6                  & 3                   & $1.49\pi$                  & 0.87                & 0.83                  & 0.92                  & 0.75        & 0.68            & 0.80            \\
    7                  & 4                   & $0.21\pi$                  & 0.56                & 0.59                  & 0.56                  & 0.15        & 0.20            & 0.27            \\
    8                  & 3                   & $0.92\pi$                  & 0.80                & 0.65                  & 0.90                  & 0.60        & 0.35            & 0.78            \\
    9                  & 4                   & $1.67\pi$                  & 0.86                & 0.84                  & 0.91                  & 0.73        & 0.63            & 0.83            \\
    10                 & 4                   & $1.16\pi$                  & 0.86                & 0.86                  & 0.92                  & 0.68        & 0.57            & 0.85            \\
    \bottomrule
  \end{tabular}
  \caption{\ac{HPOD} modes and phase shifts for the best match of the first ten \ac{POD} modes, and the corresponding $\mc{R}_{k,j}$ values.
  Superscripts $^u$ and $^v$ denote the streamwise and transverse components, respectively.}\label{tab:R_HPOD,POD}
\end{table}

\begin{figure}
  \centering
  \includegraphics[width=5.3in]{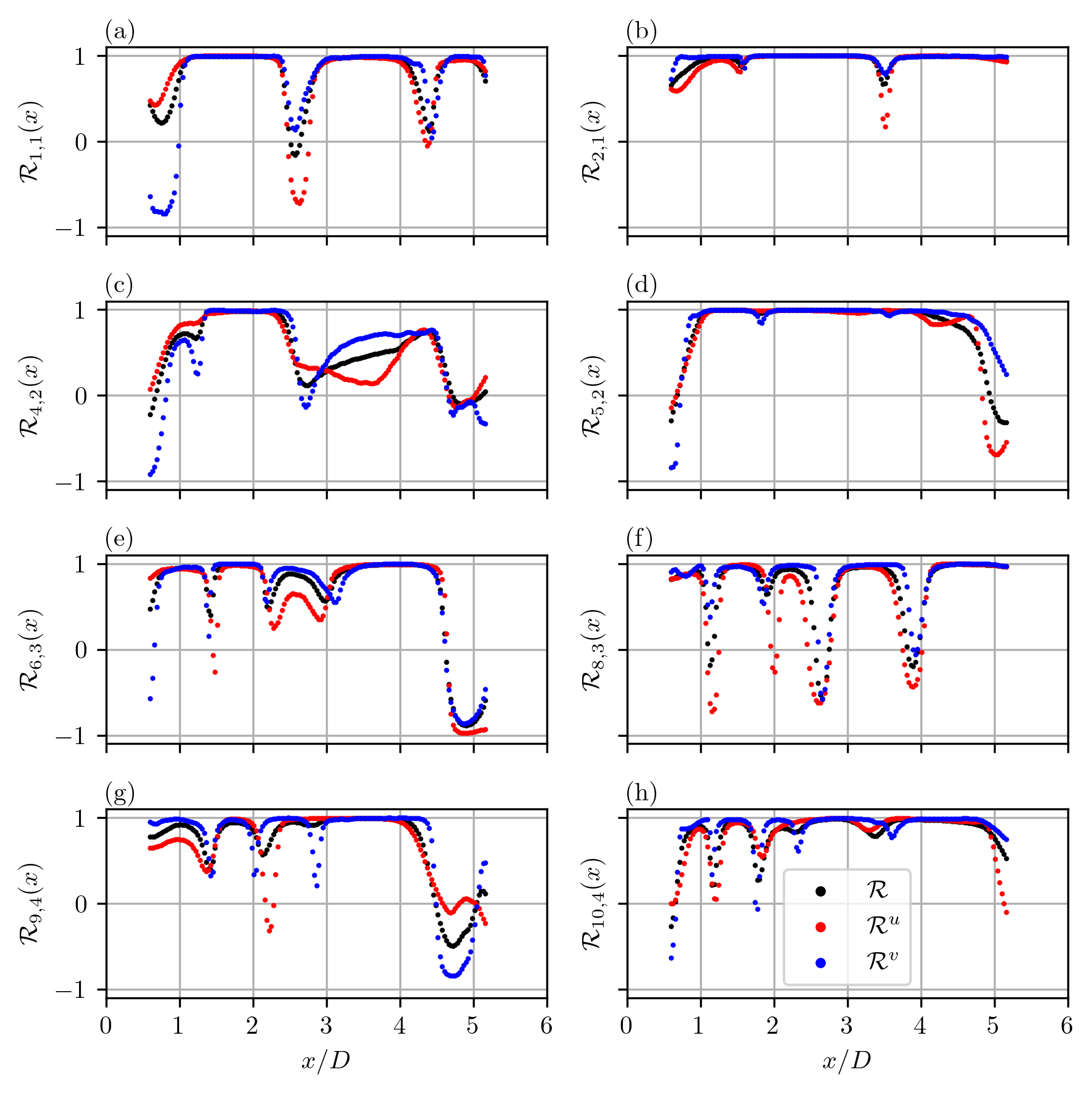}
  \caption{Correlation coefficient between each paired \ac{POD} mode and the corresponding \ac{HPOD} mode at the best match phase angle as a function of streamwise position.
  Superscripts $^u$ and $^v$ denote the streamwise and transverse components, respectively.}\label{fig:Rx_POD_HPOD}
\end{figure}

The correlation coefficient between the $j^{th}$ \ac{POD} mode and the $k^{th}$ \ac{HPOD} mode as a function of streamwise position is given by
\begin{equation}\label{eq:Rxkj}
  \mc{R}_{k,j}(x,\Delta\phi^a_{k,j}) = \frac{\cov \big( \psi^a_k(x,\Delta\phi^a_{k,j}),\psi_j(x) \big)}{\sqrt{\var \big( \psi^a_k(x,\Delta\phi^a_{k,j}) \big) \var \big( \psi_j(x) \big)}},
\end{equation}
and are shown for the matched \ac{HPOD} modes and \ac{POD} modes in figure~\ref{fig:Rx_POD_HPOD}.

The first and second \ac{POD} modes exhibit the strongest correspondence to the first \ac{HPOD} mode, with correlation coefficients of 0.88 and 0.98, and $R_{k,j}^2$ values of 0.74 and 0.95, respectively.
The corresponding phase angles are $1.64\pi$ and $1.15\pi$, which are approximately $\pi/2$ apart and consistent with the separation of the real and imaginary parts of the \ac{HPOD} modes.
This suggests that the relationship between these \ac{POD} modes mirrors that of the real and imaginary components of the first \ac{HPOD} mode.
The correlation coefficient between these \ac{POD} and \ac{HPOD} modes is consistently close to unity, with troughs in the correlation occurring due to the values of the modes approaching zero at $x/D=2.6$ and $x/D=4.4$, as shown in figure~\ref{fig:Rx_POD_HPOD}a.
Additional reductions occur near the domain boundaries, with the trough at the upstream end being significantly deeper.
While the correlation of the combined and streamwise components decreases to approximately 0.5, the correlation of the transverse component approaches -1.
This indicates that the transverse components of the first \ac{POD} mode and the first \ac{HPOD} mode have opposite signs close to the sphere.
This discrepancy likely results from a combination of spectral leakage at the domain edge and the difference in length of the structures in the \ac{HPOD} mode and \ac{POD} modes, exemplified by figures~\ref{fig:HPOD1_POD1,2}c(i) and~\ref{fig:HPOD1_POD1,2}d(i).
The reduction in correlation at the end of the domain is less severe, and most likely reflects the attenuation of the Hann window.
Similarly, the correlation coefficient of the second \ac{POD} mode with the first \ac{HPOD} mode is consistently high throughout the domain, with small dips in the correlation coefficient occurring at $x/D=1.55$ and $x/D=3.5$, as shown in figure~\ref{fig:Rx_POD_HPOD}b.
The spacing between drops in the correlation coefficient between the first \ac{HPOD} mode and both the first and second \ac{POD} modes is approximately $1.9D$, suggesting that these modes correspond to the same propagating structure, with a wavelength of approximately $3.8D$.

The fourth and fifth \ac{POD} modes are the best matches to the second \ac{HPOD} mode, with correlation coefficients of 0.86 and 0.96, and $R_{k,j}^2$ values of 0.68 and 0.84, respectively.
The corresponding phase angles are $1.71\pi$ and $1.21\pi$, which are consistent with the expected $\pi/2$ offset.
The correlation of the fourth \ac{POD} mode with the second \ac{HPOD} mode at the corresponding phase offset, shown in figure~\ref{fig:Rx_POD_HPOD}c, is close to unity between $x/D = 1.25$ and $x/D = 2.4$, after which it drops before increasing again from approximately $x/D = 2.7$.
The downstream increase never reaches unity because the fourth \ac{POD} mode is asymmetric about the centreline, as shown in figure~\ref{fig:POD_Modes}d, whereas the second \ac{HPOD} mode is symmetric, as shown in figures~\ref{fig:HPOD_Modes}c--d.
The correlation of the fifth \ac{POD} mode with the second \ac{HPOD} mode at the corresponding phase offset, shown in figure~\ref{fig:Rx_POD_HPOD}d, is consistently high across the measurement domain, except towards the edges of the domain, where the Hann window attenuates the intensity of the \ac{HPOD} modes.
This behaviour reflects the greater symmetry of the fifth \ac{POD} mode, shown in figure~\ref{fig:POD_Modes}e, compared to the fourth \ac{POD} mode, shown in figure~\ref{fig:POD_Modes}d.

The sixth and eighth \ac{POD} modes are the best matches to the third \ac{HPOD} mode, with correlation coefficients of 0.87 and 0.80, and $R_{k,j}^2$ values of 0.75 and 0.60, respectively.
The separation of the corresponding phase angles, $1.49\pi$ and $0.92\pi$, is slightly greater than $\pi/2$.
The correlation between the sixth \ac{POD} mode and the third \ac{HPOD} mode at the corresponding phase offset, shown in figure~\ref{fig:Rx_POD_HPOD}e, remains close to unity across the measurement domain with regularly spaced troughs where the mode values approach zero up to $x/D = 3$.
Downstream of this point, the correlation remains at unity for a longer streamwise distance before decreasing towards the domain edge.
This extended region of high correlation corresponds to the increased size of the structures in the sixth \ac{POD} mode at this streamwise location, shown in figure~\ref{fig:POD_Modes}f.
The substantial decrease in correlation towards the downstream end reflects the reduced size of the structures in the \ac{HPOD} mode, and the subsequent regions of opposing sign which are visible at the end of the domain in figure~\ref{fig:HPOD_Modes}f(i).
The correlation between the eighth \ac{POD} mode and the third \ac{HPOD} mode at the corresponding phase offset, shown in figure~\ref{fig:Rx_POD_HPOD}f, is consistently high across the measurement domain, with periodic troughs which increase in wavelength downstream, consistent with the structures in the eighth \ac{POD} mode, shown in figure~\ref{fig:POD_Modes}h(i).

The ninth and tenth \ac{POD} modes are the best matches to the fourth \ac{HPOD} mode at phase angles of $1.67\pi$ and $1.16\pi$, respectively.
The corresponding correlation coefficients are both 0.86, while the $R_{k,j}^2$ values are 0.73 and 0.68, respectively.
The correlation between the ninth \ac{POD} mode and the fourth \ac{HPOD} mode at the corresponding phase offset, shown in figure~\ref{fig:Rx_POD_HPOD}g, is consistently high across the measurement domain, with troughs where the mode values approach zero.
Additionally, the correlation drops towards the end of the measurement domain, with the streamwise component approaching zero at the end of the domain and the transverse component changing sign, as shown in figures~\ref{fig:POD_Modes}i(i) and~\ref{fig:POD_Modes}i(ii), respectively.
The correlation between the tenth \ac{POD} mode and the fourth \ac{HPOD} mode at the corresponding phase offset, shown in figure~\ref{fig:Rx_POD_HPOD}h, is also consistently high across the measurement domain.
The troughs in the correlation are more widely spaced further downstream, consistent with the increasing wavelength of the structures in the tenth \ac{POD} mode, shown in figure~\ref{fig:POD_Modes}j.
The correlation also drops towards the ends of the measurement domain, with the transverse component changing sign at the beginning of the domain.
The correlation between the streamwise components becomes negative at the end of the domain, reflecting structures of opposite sign appearing in the \ac{HPOD} modes, due to the spectral leakage visible in figure~\ref{fig:HPOD_Modes}g.

\begin{figure}
  \centering
  \includegraphics[width=3.5in]{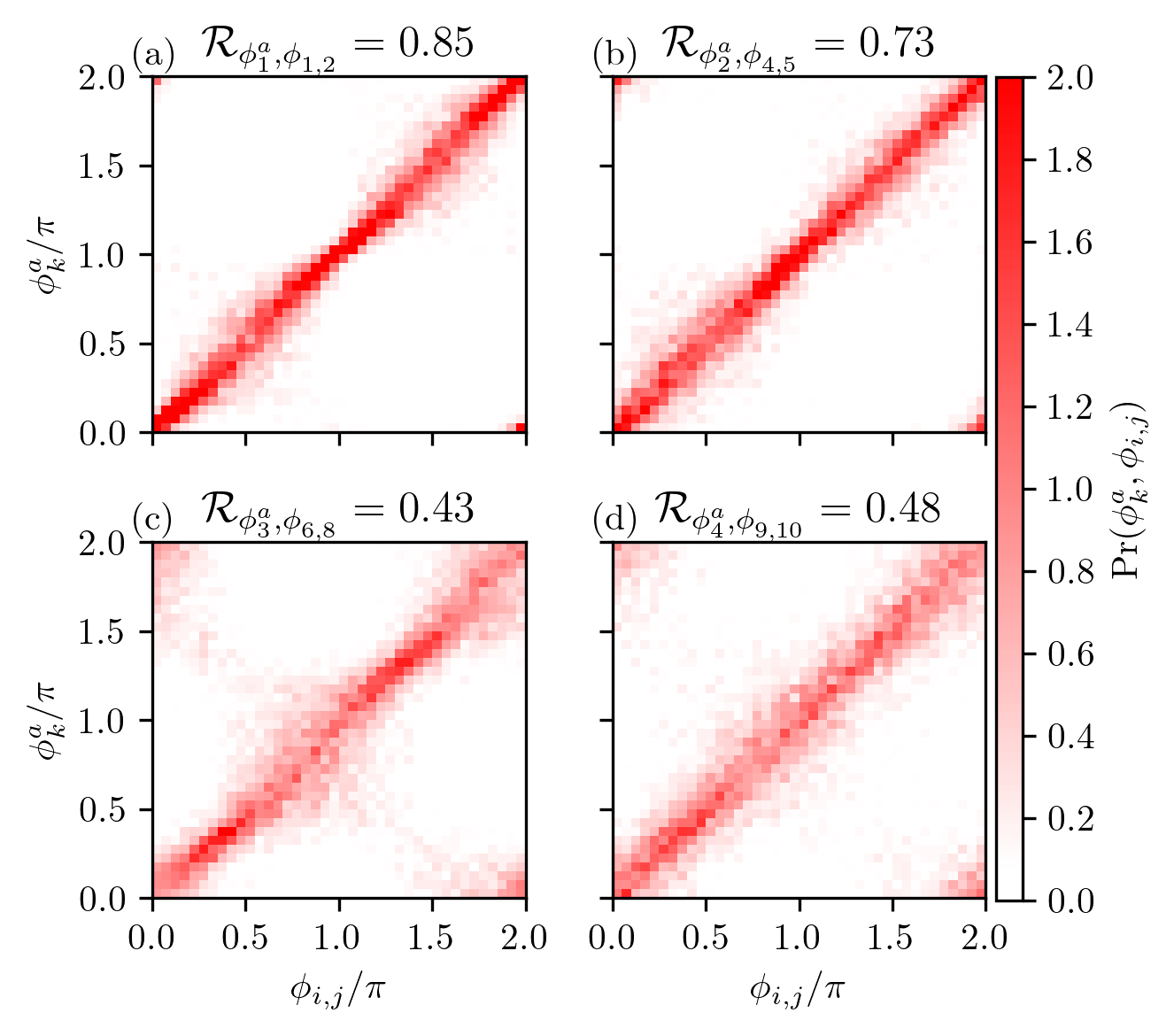}
  \caption{Joint probability density function of the phase angles of (a) the first \ac{HPOD} mode and the first and second \ac{POD} modes, (b)  the second \ac{HPOD} mode and the fourth and fifth \ac{POD} modes, (c) the third \ac{HPOD} mode and the sixth and eighth \ac{POD} modes, and (d) the fourth \ac{HPOD} mode and the ninth and tenth \ac{POD} modes.
  Phase angles of the \ac{HPOD} modes are relative to the phase offset to the first \ac{POD} mode of the corresponding pair.}\label{fig:Rphi}
\end{figure}

The $R^2_{k,\wt{(i,j)}}$ values of the $k^{th}$ \ac{HPOD} mode and the phase average of the $i^{th}$ and $j^{th}$ \ac{POD} modes are defined as
\begin{equation}\label{eq:R2_PhsAvg}
  R^2_{k,\wt{(i,j)}} = 1 - \frac{\sum_{\bs{x},\phi} {\Big(\psi^a_k(\bs{x},\phi + \Delta \phi_{k,i}^a) - \wt{\psi_{i,j}}(\bs{x}, \phi) \Big)}^2}{\sum_{\bs{x},\phi} {\Big( \wt{\psi_{i,j}}(\bs{x}, \phi) \Big)}^2},
\end{equation}
where $\bs{x}$ represents the spatial coordinates of the modes, $\Delta\phi_{k,i}$ is the phase offset between the $k^{th}$ \ac{HPOD} mode and the $i^{th}$ \ac{POD} mode, $\wt{\psi_{i,j}}$ is the phase average of the $i^{th}$ and $j^{th}$ \ac{POD} modes, and $\phi$ represents the phase angles from 0 to 2$\pi$ in 0.01$\pi$ increments.
The $R^2_{k,\wt{(i,j)}}$ value of the combined streamwise and transverse components as well as the individual components, denoted by $^u$ and $^v$, respectively, are summarised in table~\ref{tab:R2_HPOD_POD}.
The joint probability density functions of the phase angle of the $k^{th}$ \ac{HPOD} mode and the phase angle of the corresponding \ac{POD} mode pair, adjusted for the phase offset, are shown in figure~\ref{fig:Rphi}.

\begin{table}
    \centering
    \begin{tabular}{ccccc}
      \toprule
      \bf{HPOD Mode $k$} & \bf{POD Modes $i,j$} & $R^2_{k,\wt{(i,j)}}$ & $(R^2_{k,\wt{(i,j)}})^u$ & $(R^2_{k,\wt{(i,j)}})^v$ \\
      \midrule
      1                  & 1,2                  & 0.83            & 0.79                & 0.85                 \\
      2                  & 4,5                  & 0.80            & 0.78                & 0.82                 \\
      3                  & 6,8                  & 0.67            & 0.53                & 0.76                 \\
      4                  & 9,10                 & 0.70            & 0.61                & 0.83                 \\
      \bottomrule
    \end{tabular}
    \caption{$R_{k,(i,j)}^2$ values of the phase-averaged \ac{HPOD} modes and the corresponding \ac{POD} modes.
    Superscripts $^u$ and $^v$ refer to the streamwise and transverse components, respectively.}\label{tab:R2_HPOD_POD}
\end{table}

\begin{figure}
  \centering
  \includegraphics[width=5.3in]{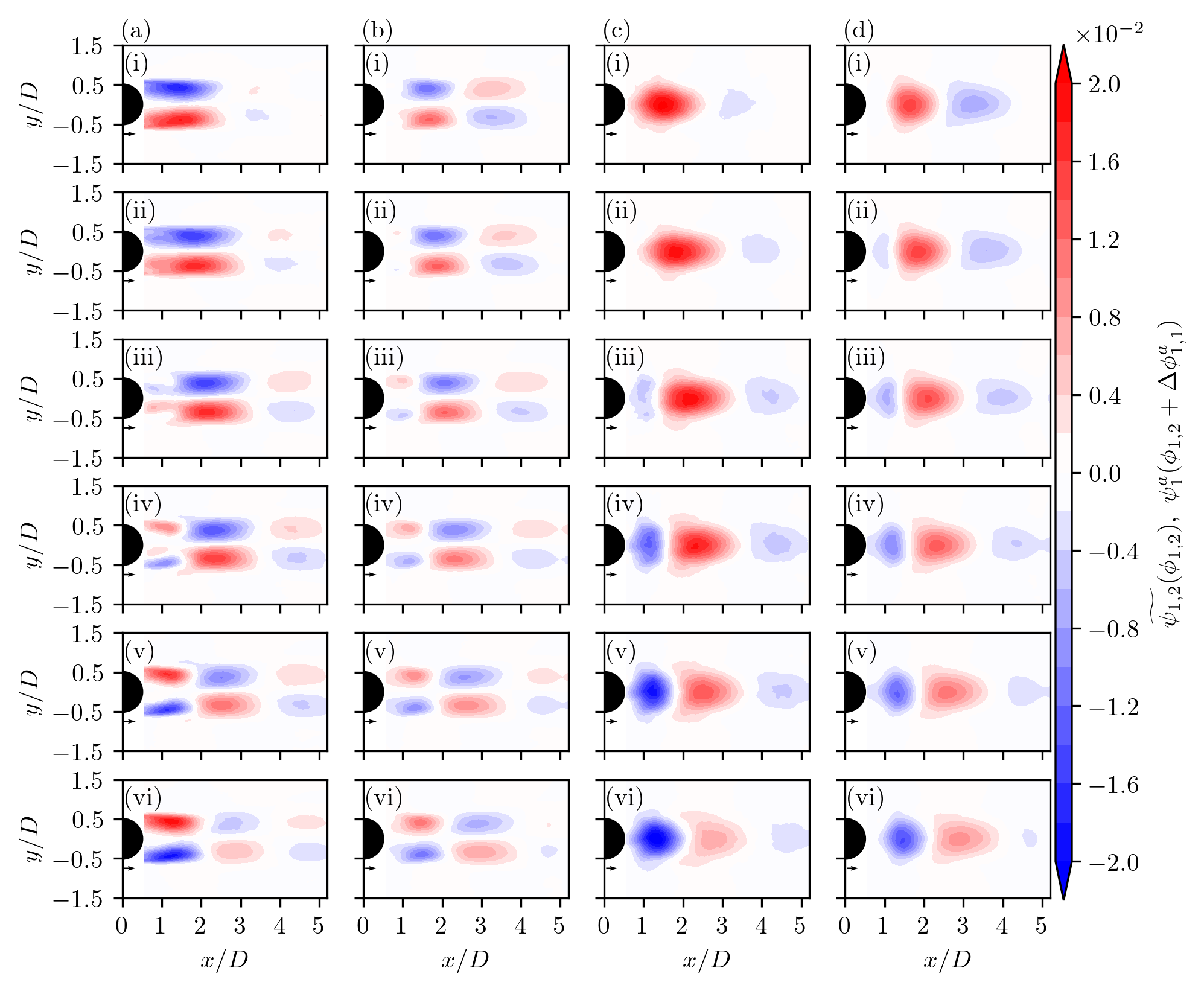}
  \caption{Phase-averaged (a) streamwise and (c) transverse components of the first and second \ac{POD} modes, and the phase-averaged (b) streamwise and (d) transverse components of the first \ac{HPOD} mode, adjusted for the phase angle offset $\Delta\phi_{1,1}$, at $\phi=$ (i) 0, (ii) $\pi/6$, (iii) $\pi/3$, (iv) $\pi/2$, (v) $2\pi/3$, and (vi) $5\pi/6$.
  See supplementary movie 1 for an animated version of this figure.}\label{fig:HPOD1_POD1,2}
\end{figure}

The streamwise component of the phase-averaged first and second \ac{POD} modes, shown in figure~\ref{fig:HPOD1_POD1,2}a, agrees well with that of the first \ac{HPOD} mode at the adjusted phase angle, shown in figure~\ref{fig:HPOD1_POD1,2}b.
The main differences in these structures result from the accentuation of propagating structures by \ac{HPOD}.
Specifically, the structures in the \ac{POD} mode exhibit greater intensity directly behind the sphere, while those further downstream are more pronounced in the \ac{HPOD} modes.
This difference is also evident in the transverse components, shown in figures~\ref{fig:HPOD1_POD1,2}c and~\ref{fig:HPOD1_POD1,2}d for the \ac{POD} modes and \ac{HPOD} modes, respectively.
Additionally, the shape of the structures in the near wake of the sphere in the \ac{POD} modes, shown in figure~\ref{fig:HPOD1_POD1,2}a(iv), differs from that in the \ac{HPOD} mode, shown in figure~\ref{fig:HPOD1_POD1,2}b(iv).
This results from the assumed periodicity of the Hilbert transform, which couples the near-wake structures with those further downstream.
Despite these differences, it is evident that the first and second \ac{POD} modes and the first \ac{HPOD} mode represent the same propagating structure.
This structure consists of planar-antisymmetric streamwise fluctuations and alternating positive and negative transverse fluctuations, consistent with wake flapping.
The similarity of the third \ac{POD} mode, shown in figure~\ref{fig:POD_Modes}c, to the first and second suggests that the flapping includes an out-of-plane component.

\begin{figure}
  \centering
  \includegraphics[width=5.3in]{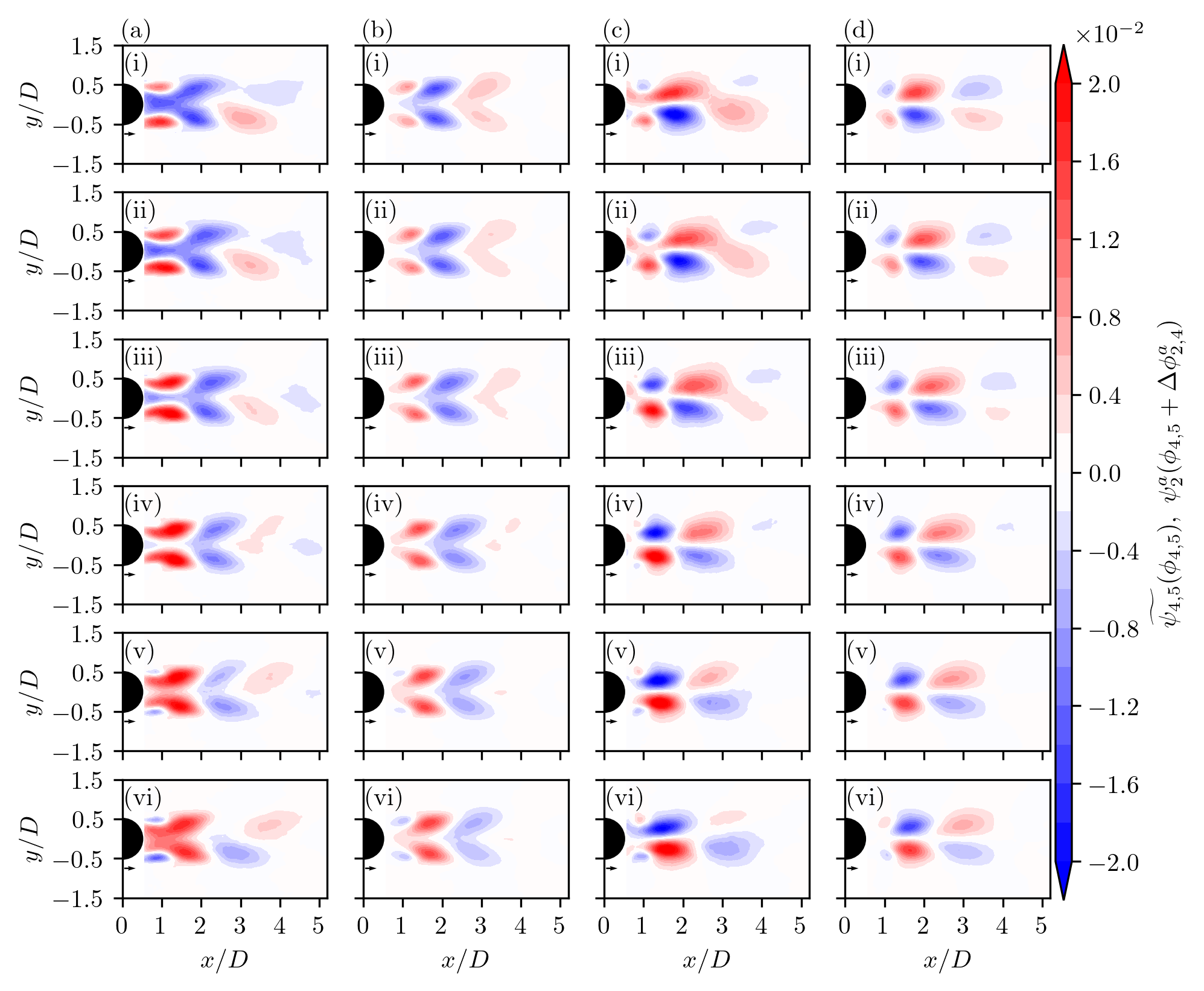}
  \caption{Phase-averaged (a) streamwise and (c) transverse components of the fourth and fifth \ac{POD} modes, and the phase-averaged (b) streamwise and (d) transverse components of the second \ac{HPOD} mode, adjusted for the phase angle offset $\Delta\phi_{2,4}$, at $\phi=$ (i) 0, (ii) $\pi/6$, (iii) $\pi/3$, (iv) $\pi/2$, (v) $2\pi/3$, and (vi) $5\pi/6$.
  See supplementary movie 2 for an animated version of this figure.}\label{fig:HPOD2_POD4,5}
\end{figure}

While the streamwise component of the phase-averaged fourth and fifth \ac{POD} modes, shown in figure~\ref{fig:HPOD2_POD4,5}a, resembles the second \ac{HPOD} mode, shown in figure~\ref{fig:HPOD2_POD4,5}b, the latter exhibits markedly greater symmetry, attributed to \ac{HPOD}'s inherent emphasis on propagating features.
This is most evident when comparing figures~\ref{fig:HPOD2_POD4,5}a\@(ii) and~\ref{fig:HPOD2_POD4,5}a\@(vi) to figures~\ref{fig:HPOD2_POD4,5}b\@(ii) and~\ref{fig:HPOD2_POD4,5}b\@(vi).
In the \ac{HPOD} case, the V-shaped structures remain symmetric downstream, whereas they become asymmetric in the \ac{POD} modes.
Additionally, the V-shaped structures extend further upstream structures close to the sphere ($x/D < 1.5$) in the \ac{POD} modes, but remain distinct in the \ac{HPOD} modes. 
The transverse component of the second \ac{HPOD} mode, shown in figure~\ref{fig:HPOD2_POD4,5}d, exhibits greater antisymmetry than the transverse components of the phase average of the fourth and fifth \ac{POD} modes, shown in figure~\ref{fig:HPOD2_POD4,5}c, particularly upstream of $x/D = 3$.
The asymmetry in the \ac{POD} modes likely arises because three-dimensionality in the flow is more pronounced in the structures\ represented by the fourth and fifth \ac{POD} modes than that represented by the first and second \ac{POD} modes.
In this case, the use of \ac{HPOD} proves advantageous for isolating the propagating structures in the flow in the presence of asymmetry in the \ac{POD} modes.
The alternating positive and negative streamwise fluctuations in these modes indicates that they correspond to a pulsating motion in the sphere's wake.

\begin{figure}
  \centering
  \includegraphics[width=5.3in]{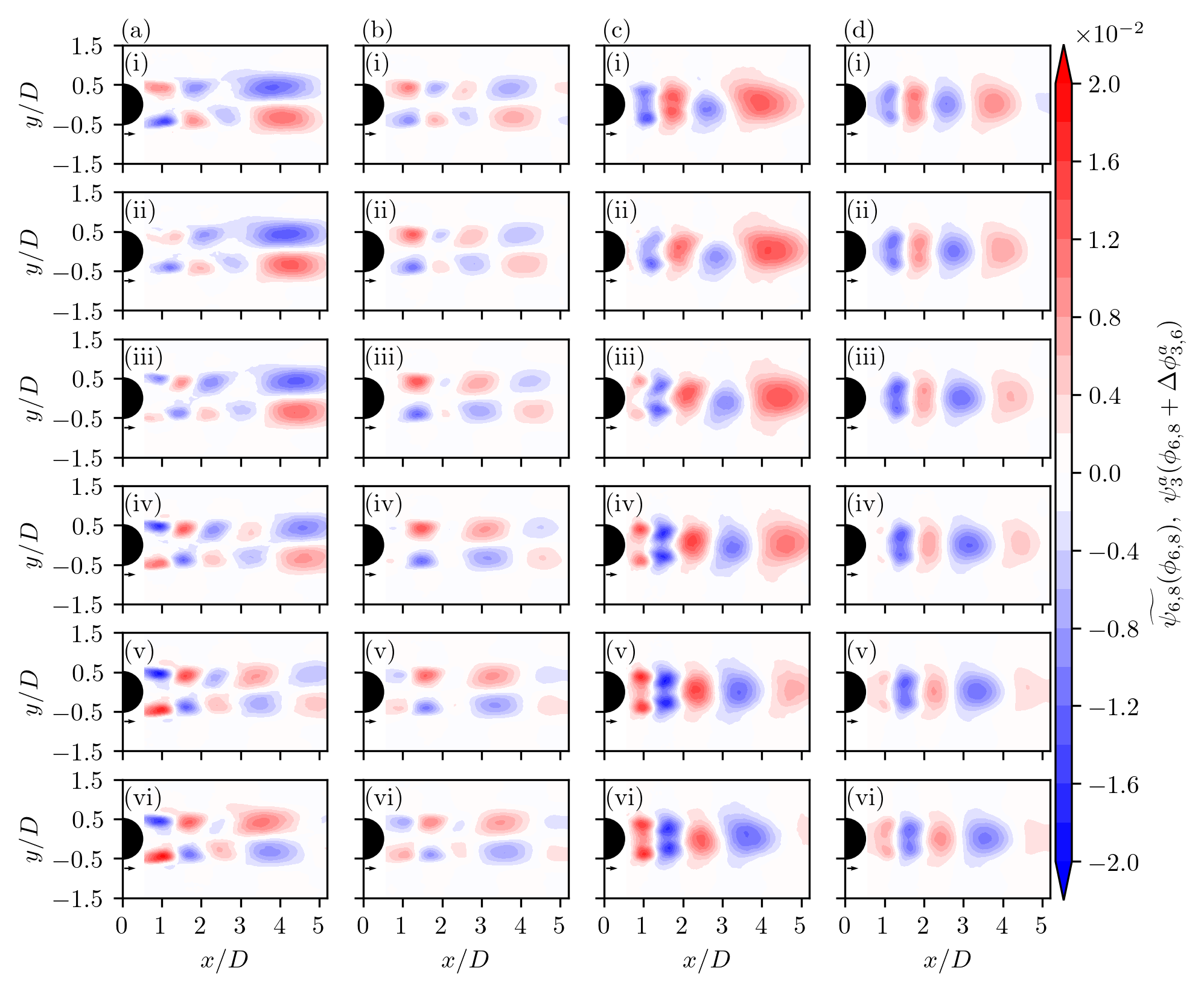}
  \caption{Phase-averaged (a) streamwise and (c) transverse components of the sixth and eighth \ac{POD} modes, and the phase-averaged (b) streamwise and (d) transverse components of the third \ac{HPOD} mode, adjusted for the phase angle offset $\Delta\phi_{3,6}$, at $\phi=$ (i) 0, (ii) $\pi/6$, (iii) $\pi/3$, (iv) $\pi/2$, (v) $2\pi/3$, and (vi) $5\pi/6$.
  See supplementary movie 3 for an animated version of this figure.}\label{fig:HPOD3_POD6,8}
\end{figure}

The streamwise component of the phase average of the sixth and eighth \ac{POD} modes, shown in figure~\ref{fig:HPOD3_POD6,8}a, is qualitatively similar to that of the third \ac{HPOD} mode, shown in figure~\ref{fig:HPOD3_POD6,8}b.
However, the alternating positive and negative structures are more distinctly defined in the \ac{HPOD} mode than the phase average of the \ac{POD} modes.
This is particularly evident when comparing figures~\ref{fig:HPOD3_POD6,8}a(i) and~\ref{fig:HPOD3_POD6,8}a(ii) to figures~\ref{fig:HPOD3_POD6,8}b(i) and~\ref{fig:HPOD3_POD6,8}b(ii), in which the positive region in the \ac{HPOD} modes around $x/D = 3$ and $y/D > 0$ is not present in the phase-averaged \ac{POD} modes.
This likely attributable to the out-of-plane motion of the structure as it propagates, while the Hilbert transform enforces periodicity.
Additionally, the intensity of the structures tends to be lower in the \ac{HPOD} modes than those in the \ac{POD} modes because the non-propagating components are relegated from the leading \ac{HPOD} modes.
The transverse component of the phase average of the sixth and eighth \ac{POD} modes, shown in figure~\ref{fig:HPOD3_POD6,8}c, exhibits transverse displacement of alternating-sign structures, most evident in figures~\ref{fig:HPOD3_POD6,8}c(i)--(iii).
This shifting is absent from the transverse component of the third \ac{HPOD} mode, shown in figure~\ref{fig:HPOD3_POD6,8}d.
The shifting of the structures in the transverse velocity of the \ac{POD} modes aligns with the three-dimensional motion of the wake and corresponds to the observed behaviour of their streamwise components.
The structure represented by these modes suggests wake flapping with a shorter wavelength than that associated with the first and second \ac{POD} modes and the first \ac{HPOD} mode.

\begin{figure}
  \centering
  \includegraphics[width=5.3in]{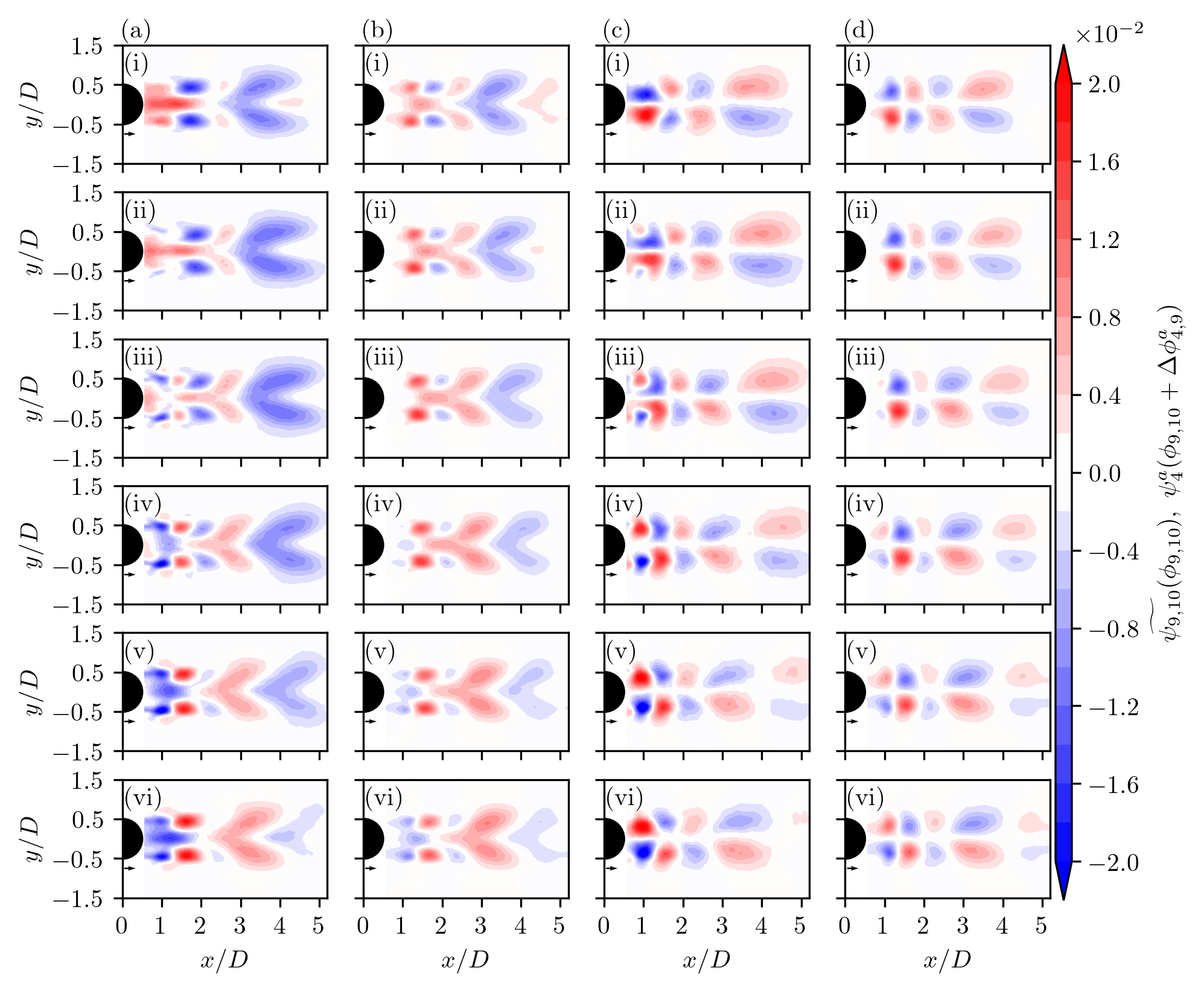}
  \caption{Phase-averaged (a) streamwise and (c) transverse components of the ninth and tenth \ac{POD} modes, and the phase-averaged (b) streamwise and (d) transverse components of the fourth \ac{HPOD} mode, adjusted for the phase angle offset $\Delta\phi_{4,9}$, at $\phi=$ (i) 0, (ii) $\pi/6$, (iii) $\pi/3$, (iv) $\pi/2$, (v) $2\pi/3$, and (vi) $5\pi/6$.
  See supplementary movie 4 for an animated version of this figure.}\label{fig:HPOD4_POD9,10}
\end{figure}

The streamwise component of the phase-averaged ninth and tenth \ac{POD} modes, shown in figure~\ref{fig:HPOD4_POD9,10}a, is qualitatively comparable to the phase average of the fourth \ac{HPOD} mode, shown in figure~\ref{fig:HPOD4_POD9,10}b, for $x/D > 2.5$.
In this domain, the streamwise components of the \ac{POD} and \ac{HPOD} modes exhibit V-shaped structures similar to those in the fourth and fifth \ac{POD} modes and the second \ac{HPOD} mode.
For $x/D < 2.5$, the structure of the \ac{POD} modes exhibits greater detail, most evident in figures~\ref{fig:HPOD4_POD9,10}a(iii) and~\ref{fig:HPOD4_POD9,10}a(iv) compared with figures~\ref{fig:HPOD4_POD9,10}b(iii) and~\ref{fig:HPOD4_POD9,10}b(iv).
The transverse components of the phase average of the ninth and tenth \ac{POD} modes and the phase average of the fourth \ac{HPOD} mode, shown in figures~\ref{fig:HPOD4_POD9,10}c and~\ref{fig:HPOD4_POD9,10}d, respectively, also exhibit qualitative similarity.
Similar to the streamwise components, the transverse component of the phase-averaged \ac{POD} modes includes features absent from the \ac{HPOD} mode in the immediate near wake, attributable to the assumed periodicity of the Hilbert transform.
This is clearly evident in figures~\ref{fig:HPOD4_POD9,10}c(ii) and~\ref{fig:HPOD4_POD9,10}c(iii) compared to figures~\ref{fig:HPOD4_POD9,10}d(ii) and~\ref{fig:HPOD4_POD9,10}d(iii) around $x/D = 1$.

The joint probability density functions of the \ac{HPOD} modes and the corresponding \ac{POD} mode pairs, shown in figure~\ref{fig:Rphi}, are highest around $\phi^a_k = \phi_{i,j}$ when adjusted for the appropriate phase offset.
This indicates that the phase relationship between each \ac{HPOD} mode and its corresponding \ac{POD} mode pair is consistent throughout the measurement domain.
The spread of the distributions varies between the different mode pairs, with the tightest clustering occurring for the first \ac{HPOD} mode and the first and second \ac{POD} modes, shown in figure~\ref{fig:Rphi}a, and the widest spread being that of the third \ac{HPOD} mode and the sixth and eighth \ac{POD} modes, shown in figure~\ref{fig:Rphi}c.
This is consistent with the $R^2_{k,\wt{(i,j)}}$ values of the mode shapes summarised in table~\ref{tab:R2_HPOD_POD}, with the highest value of 0.83 occurring for the first \ac{HPOD} mode and the first and second \ac{POD} modes, and the lowest value of 0.67 occurring for the third \ac{HPOD} mode and the sixth and eighth \ac{POD} modes.

\subsection{Pairing POD Modes using the Hilbert Transform}

While \ac{HPOD} modes can be used to identify matched pairs of \ac{POD} modes, this process requires performing both decompositions.
Because the Hilbert transform can be applied in the streamwise direction, it can be applied directly to the \ac{POD} modes, despite their lack of temporal information.
The result is a $\mp \pi/2$ phase shift of the fundamental Fourier components of each \ac{POD} mode, analogous to the downstream propagation of the mode.
Thus, the streamwise Hilbert transform of a \ac{POD} mode is analogous to a \ac{POD} mode that represents a $\mp \pi/2$ phase shift of the same streamwise periodic structure.

In order to determine matched pairs of \ac{POD} modes using the Hilbert transform, the Hilbert transform of the $k^{th}$ \ac{POD} mode in the streamwise direction $\mc{H}_x[\psi_k(\bs{x})]$ is used to define the analytic signal of $\psi_k(\bs{x})$
\begin{equation}\label{eq:psiAnalytic}
  {( \psi_k )}^a(\bs{x}) = \psi_k(\bs{x}) + i \mc{H}_x \big[ \psi_k (\bs{x}) \big].
\end{equation}
The shape of the analytic signal of the $k^{th}$ \ac{POD} mode at the phase angle $\phi_k$ is given by
\begin{equation}\label{eq:phiAnalytic}
  \big( \psi_k \big)^a(\bs{x},\phi_k)= \cos(\phi_k) \psi_k(\bs{x}) + \sin(\phi_k)\mc{H}_x[\psi_k(\bs{x})],
\end{equation}
and the phase angle at which the $k^{th}$ \ac{POD} mode best matches the $j^{th}$ \ac{POD} mode is given by
\begin{equation}\label{eq:pairAnalytic}
  \begin{bmatrix}
    \cos(\Delta \phi_{k,1}) & \sin(\Delta \phi_{k,1}) \\
    \vdots                  & \vdots                  \\
    \cos(\Delta \phi_{k,j}) & \sin(\Delta \phi_{k,j}) \\
    \vdots                  & \vdots                  \\
    \cos(\Delta \phi_{k,K}) & \sin(\Delta \phi_{k,K}) \\
  \end{bmatrix}^T
  =
  \left( \begin{bmatrix}
    \psi_k(\bs{x}) \\ \mc{H}_x[\psi_k(\bs{x})] \\
    \end{bmatrix}
    \begin{bmatrix}
    \psi_k(\bs{x}) \\ \mc{H}_x[\psi_k(\bs{x})] \\
    \end{bmatrix}^T \right)^{-1}
    \begin{bmatrix}
    \psi_k(\bs{x}) \\ \mc{H}_x[\psi_k(\bs{x})]
    \end{bmatrix}
    \begin{bmatrix}
    \psi_1(\bs{x}) \\ \vdots \\ \psi_j(\bs{x}) \\ \vdots \\ \psi_K(\bs{x})\\
  \end{bmatrix}^T,
\end{equation}
and
\begin{equation}\label{eq:Detlaphikj}
  \Delta \phi_{k,j} = \tan^{-1} \frac{\sin(\Delta \phi_{k,j})}{\cos(\Delta \phi_{k,j})}.
\end{equation}
The corresponding correlation coefficient is
\begin{equation}\label{eq:R_H[POD]_POD}
  \mc{R}_{k^a,j}(\Delta \phi_{k,j}) = \frac{\cov \big( (\psi_k)^a(\bs{x},\Delta\phi_{k,j}) , \psi_j(\bs{x}) \big)}{\sqrt{\var \big( (\psi_k)^a(\bs{x},\Delta\phi_{k,j}) \big)\var \big( \psi_j(\bs{x}) \big)}}.
\end{equation}
The $R^2_{k^a,\wt{(k,j)}}$ value of the phase average of the $k^{th}$ and $j^{th}$ \ac{POD} modes compared to the phase average of the analytic signal of the $k^{th}$ \ac{POD} mode is given by
\begin{equation}\label{eq:R2_ka,kj}
  R^2_{k^a,\wt{(k,j)}} = 1 - \frac{\sum_{\bs{x},\phi} \big( (\psi_k)^a(\bs{x},\phi) - \wt{\psi}_{k,j}(\bs{x},\phi) \big)^2}{\sum_{\bs{x},\phi} \big( \wt{\psi}_{k,j}(\bs{x},\phi) \big)^2},
\end{equation}
where $\bs{x}$ denotes the spatial coordinates, $(\psi_k)^a(\bs{x},\phi)$ is the analytic signal of the $k^{th}$ \ac{POD} mode and $\wt{\psi}_{k,j}(\bs{x},\phi)$ is the phase average of the $k^{th}$ and $j^{th}$ \ac{POD} modes at phase angle $\phi$, which was determined using phase angles $\phi=[0,2\pi]$ in increments of $0.01\pi$.
The \ac{POD} mode which best matches the analytic signal of each of the first ten \ac{POD} modes, is given in table~\ref{tab:R_H[POD],POD}, along with the corresponding correlation coefficient, and the $R^2_{k^a,\wt{(k,j)}}$ of the phase averages.
Both the correlation coefficients and $R^2_{k^a,\wt{(k,j)}}$ values were calculated for the combined streamwise and transverse components of the modes, as well as the individual components.

\begin{table}
\centering
  \begin{tabular}{c c r c c c c c c}
    \toprule
    \bf{POD Mode $k$} & \bf{POD Mode $j$} & $\Delta \phi_{k,j}$ & $\mc{R}_{k^a,j}$ & $\mc{R}_{k^a,j}^u$ & $\mc{R}_{k^a,j}^v$ & $R^2_{k^a,\wt{(k,j)}}$ & $(R^2_{k^a,\wt{(k,j)}})^u$  & $(R^2_{k^a,\wt{(k,j)}})^v$ \\
    \midrule
    1                      &  2                            & $-0.47\pi$          & 0.85             & 0.82               & 0.89               & 0.86                   & 0.82                        & 0.88                       \\
    2                      &  1                            & $ 0.46\pi$          & 0.78             & 0.75               & 0.82               & 0.78                   & 0.74                        & 0.81                       \\
    3                      &  2                            & $ 0.47\pi$          & 0.77             & 0.75               & 0.79               & 0.79                   & 0.77                        & 0.80                       \\
    4                      &  5                            & $-0.51\pi$          & 0.86             & 0.84               & 0.89               & 0.87                   & 0.84                        & 0.89                       \\
    5                      &  4                            & $ 0.51\pi$          & 0.79             & 0.76               & 0.82               & 0.79                   & 0.75                        & 0.82                       \\
    6                      &  8                            & $-0.49\pi$          & 0.81             & 0.73               & 0.86               & 0.83                   & 0.77                        & 0.87                       \\
    7                      &  9                            & $-0.47\pi$          & 0.59             & 0.61               & 0.58               & 0.67                   & 0.67                        & 0.67                       \\
    8                      &  6                            & $ 0.48\pi$          & 0.73             & 0.63               & 0.82               & 0.73                   & 0.61                        & 0.80                       \\
    9                      & 10                            & $-0.46\pi$          & 0.75             & 0.71               & 0.82               & 0.75                   & 0.70                        & 0.83                       \\
    10                     &  9                            & $ 0.45\pi$          & 0.71             & 0.65               & 0.80               & 0.71                   & 0.65                        & 0.80                       \\
  \bottomrule
  \end{tabular}
  \caption{Paired \ac{POD} modes using the analytic signal of the $k^{th}$ \ac{POD} mode, and the corresponding phase angle, correlation coefficient and $R^2_{k^a,\wt{(k,j)}}$ of the phase averages.
  Superscripts $^u$ and $^v$ refer to the streamwise and transverse components, respectively.}\label{tab:R_H[POD],POD}
\end{table}

For each of the first ten \ac{POD} modes, the paired \ac{POD} mode corresponds to the analytic signal at a phase angle of $\mp\pi/2$.
Consequently, the $j^{th}$ \ac{POD} mode best matches the Hilbert transform of the corresponding $k^{th}$ \ac{POD} mode.
A pair of \ac{POD} modes representing a structure propagating in the streamwise direction therefore requires that the \ac{POD} modes be a mutual best match with phase angle shifts of $\pm 0.5\pi$.
Based on this criterion, the \ac{POD} mode pairs identified using the Hilbert transform are the first and second \ac{POD} modes, the fourth and fifth \ac{POD} modes, the sixth and eighth \ac{POD} modes, and the ninth and tenth \ac{POD} modes.
These are the same pairings as those identified using the \ac{HPOD} modes.

\begin{figure}
  \centering
  \includegraphics[width=5.3in]{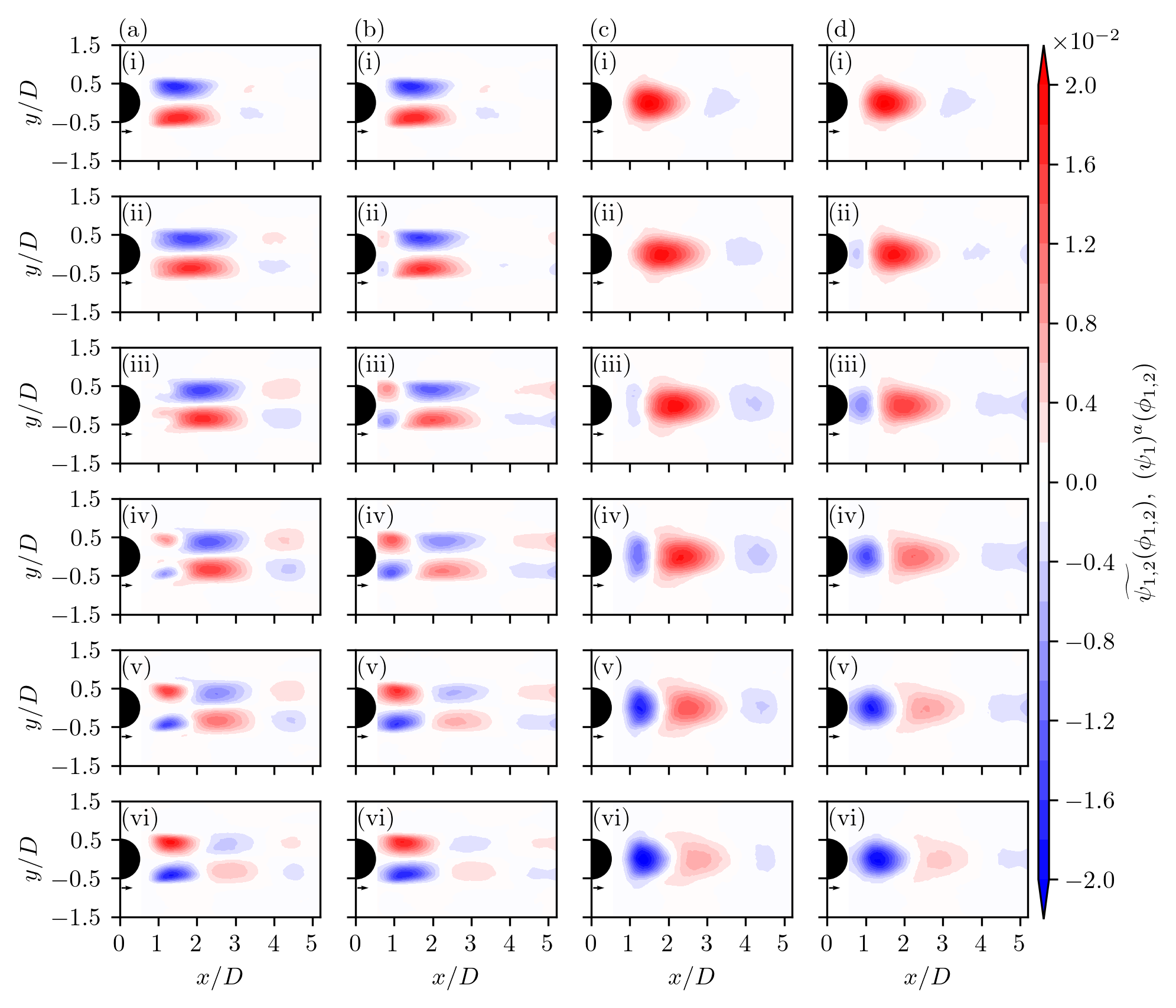}
  \caption{Phase-averaged (a) streamwise and (c) transverse components of the first and second \ac{POD} modes and the phase-averaged (b) streamwise and (d) transverse components of the phase average of the analytic signal of the first \ac{POD} mode at $\phi=$ (i) 0, (ii) $\pi/6$, (iii) $\pi/3$, (iv) $\pi/2$, (v) $2\pi/3$, and (vi) $5\pi/6$.
  See supplementary movie 5 for an animated version of this figure.}\label{fig:H[POD1]_POD1,2}
\end{figure}

The streamwise component of the analytic signal of the first \ac{POD} mode, shown in figure~\ref{fig:H[POD1]_POD1,2}b, closely matches the phase average of the first two \ac{POD} modes, shown in figure~\ref{fig:H[POD1]_POD1,2}a.
Both exhibit extended regions of alternating sign propagating downstream from the sphere. The primary difference occurs for $x/D < 2$, where the structures in the \ac{POD} modes are seen coming around the sphere, as illustrated in figure~\ref{fig:HPOD1_POD1,2}a(iv).
Conversely, the assumed periodicity of the Hilbert transform couples these structures with those further downstream in the \ac{HPOD} mode, as shown in figure~\ref{fig:H[POD1]_POD1,2}b(iv).
This discrepancy arises from applying the Hilbert transform to the first \ac{POD} mode, as these structures exist in the second \ac{POD} mode but not in the first.
Consequently, the $R^2_{k^a,\wt{(k,j)}}$ values differ: 0.89 for the analytic signal of the first mode when compared to the second mode, and 0.76 for the reverse comparison.
This variation arises because non-propagating components in one mode are absent from the other but are propagated by the Hilbert transform.
The variation between the transverse components of the phase-averaged first and second \ac{POD} modes, shown in figure~\ref{fig:H[POD1]_POD1,2}c, and the analytic signal of the first \ac{POD} mode, shown in figure~\ref{fig:H[POD1]_POD1,2}d, is considerably smaller than that observed in the streamwise components.  
Both exhibit symmetric regions of alternating sign that elongate and weaken in intensity as they propagate downstream.
This behaviour is consistent with the transverse fluctuations generated by the flow around the sphere decaying as they travel downstream. 

\begin{figure}
  \centering
  \includegraphics[width=5.3in]{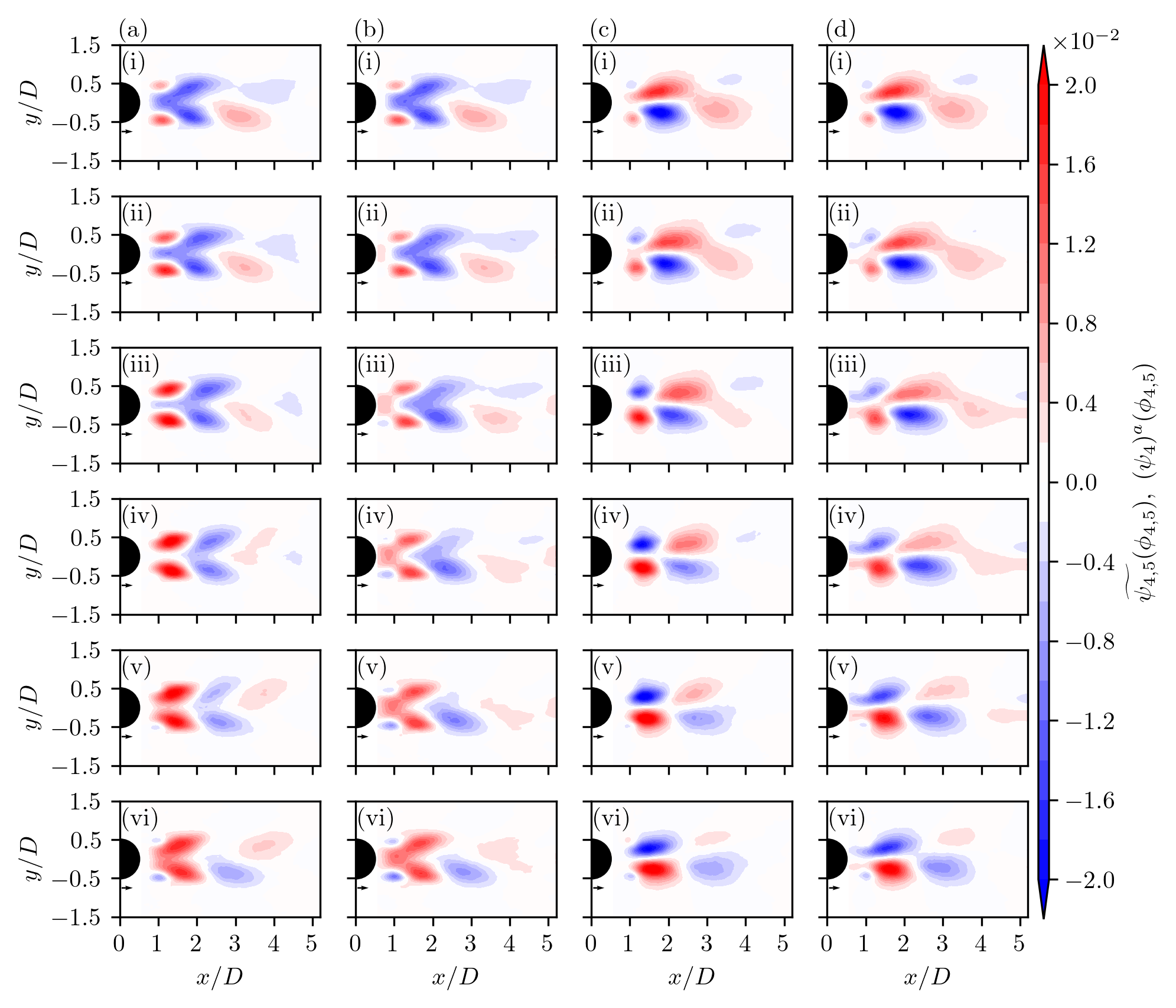}
  \caption{Phase-averaged (a) streamwise and (c) transverse components of the fourth and fifth \ac{POD} modes and the phase-averaged (b) streamwise and (d) transverse components of the phase average of the analytic signal of the fourth \ac{POD} mode at $\phi=$ (i) 0, (ii) $\pi/6$, (iii) $\pi/3$, (iv) $\pi/2$, (v) $2\pi/3$, and (vi) $5\pi/6$.
  See supplementary movie 6 for an animated version of this figure.}\label{fig:H[POD4]_POD4,5}
\end{figure}

The streamwise component of the analytic signal of the fourth \ac{POD} mode, shown in figure~\ref{fig:H[POD4]_POD4,5}b, retains more of the asymmetry of the phase-averaged fourth and fifth \ac{POD} modes, shown in figure~\ref{fig:H[POD4]_POD4,5}a, than is observed in the second \ac{HPOD} mode, shown in figure~\ref{fig:HPOD2_POD4,5}b.
The transverse component of the analytic signal of the fourth \ac{POD} mode, shown in figure~\ref{fig:H[POD4]_POD4,5}d, also retains the asymmetry present in the phase average of the fourth and fifth \ac{POD} modes, shown in figure~\ref{fig:H[POD4]_POD4,5}c.
Although the bias of the Hilbert transform towards propagating structures is less pronounced in the analytic signal of the fourth \ac{POD} mode than in the second \ac{HPOD} mode, shown in figures~\ref{fig:HPOD2_POD4,5}b and~\ref{fig:HPOD2_POD4,5}d, it remains evident, particularly in figures~\ref{fig:H[POD4]_POD4,5}b(iii) and~\ref{fig:H[POD4]_POD4,5}d(iv).

\begin{figure}
  \centering
  \includegraphics[width=5.3in]{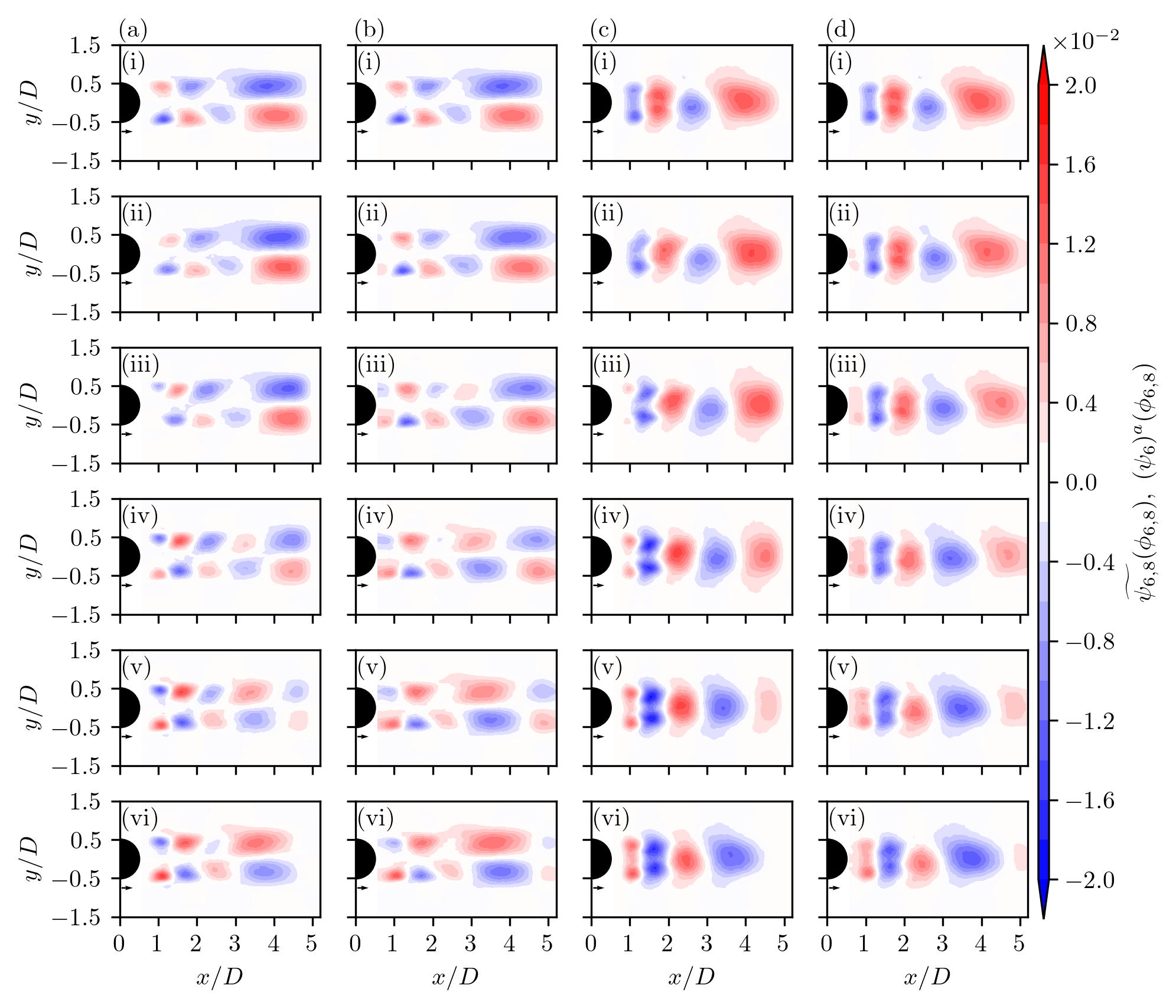}
  \caption{Phase-averaged (a) streamwise and (c) transverse components of the sixth and eighth \ac{POD} modes and the phase-averaged (b) streamwise and (d) transverse components of the phase average of the analytic signal of the sixth \ac{POD} mode at $\phi=$ (i) 0, (ii) $\pi/6$, (iii) $\pi/3$, (iv) $\pi/2$, (v) $2\pi/3$, and (vi) $5\pi/6$.
  See supplementary movie 7 for an animated version of this figure.}\label{fig:H[POD6]_POD6,8}
\end{figure}

While the analytic signal of the sixth \ac{POD} mode exhibits more symmetry than the phase average of the sixth and eighth \ac{POD} modes, its influence on the turbulent structures is less pronounced than that observed in the third \ac{HPOD}.
The analytic signal of the sixth \ac{POD} mode introduces the negative region at $x/D = 2.5$ in the streamwise component shown in figure~\ref{fig:H[POD6]_POD6,8}a(v), similar to that observed in the third \ac{HPOD} mode shown in figure~\ref{fig:HPOD3_POD6,8}b(v), which is not present in the phase average of the sixth and eighth \ac{POD} modes, shown in figure~\ref{fig:H[POD6]_POD6,8}b(v).
The analytic signal of the sixth \ac{POD} mode also retains more of the transverse shifting in the structures in the transverse velocity, as seen in figure~\ref{fig:H[POD6]_POD6,8}d, than the third \ac{HPOD} mode, shown in figure~\ref{fig:HPOD3_POD6,8}d.
Additionally, the intensity of the structures further downstream is preserved in the analytic signal of the sixth \ac{POD} mode, as illustrated in figure~\ref{fig:H[POD6]_POD6,8}b(i), where the same region exhibits significantly reduced intensity in the third \ac{HPOD} mode, shown in figure~\ref{fig:HPOD3_POD6,8}b(i).

\begin{figure}
  \centering
  \includegraphics[width=5.3in]{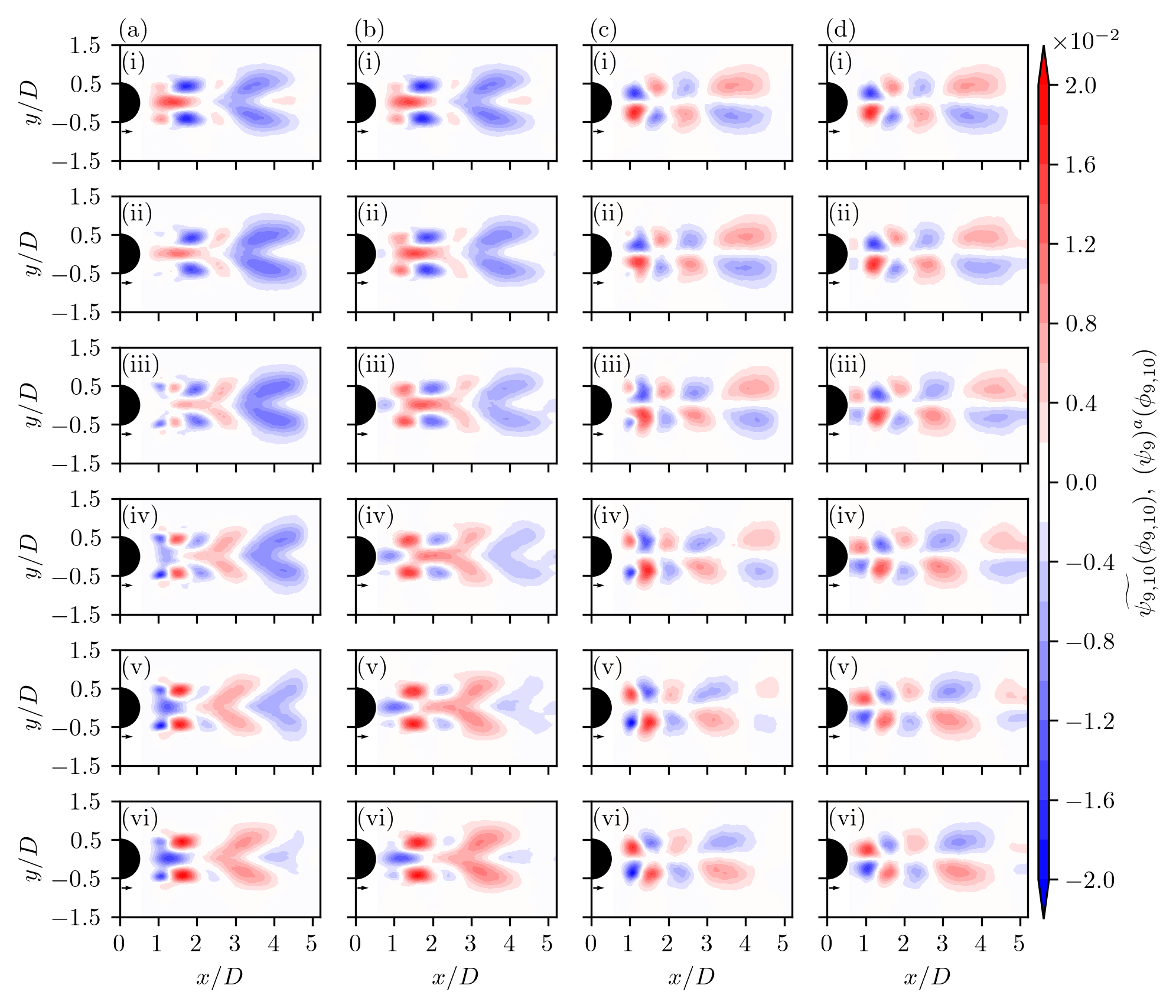}
  \caption{Phase-averaged (a) streamwise and (c) transverse components of the ninth and tenth \ac{POD} modes and the phase-averaged (b) streamwise and (d) transverse components of the phase average of the analytic signal of the ninth \ac{POD} mode at $\phi=$ (i) 0, (ii) $\pi/6$, (iii) $\pi/3$, (iv) $\pi/2$, (v) $2\pi/3$, and (vi) $5\pi/6$.
  See supplementary movie 8 for an animated version of this figure.}\label{fig:H[POD9]_POD9,10}
\end{figure}

While there is a significant difference between the streamwise components of the analytic signal of the ninth \ac{POD} mode, shown in figure~\ref{fig:H[POD9]_POD9,10}b, and that of the phase average of the ninth and tenth \ac{POD} modes, shown in figure~\ref{fig:H[POD9]_POD9,10}a, for $x/D < 2$, they become more similar further downstream of the sphere.
The structures in the phase average of the ninth and tenth modes are smaller in scale for $x/D < 2$ than they are for $x/D > 2$, which is consistent with the breakdown of smaller-scale fluctuations in the near wake as they travel downstream while the larger-scale structures persist.
Consequently, these structures are less significant to the Hilbert transform and therefore attenuated in the analytic signal.
The differences between the transverse components of the analytic signal of the ninth \ac{POD} mode, shown in figure~\ref{fig:H[POD9]_POD9,10}d, and the phase average of the ninth and tenth \ac{POD} modes, shown in figure~\ref{fig:H[POD9]_POD9,10}c, are less pronounced.
However, some variation remains in the upstream structures at $x/D > 1.5$, as shown in figure~\ref{fig:H[POD9]_POD9,10}c(ii) and~\ref{fig:H[POD9]_POD9,10}d(ii), and figures~\ref{fig:H[POD9]_POD9,10}c(iii) and~\ref{fig:H[POD9]_POD9,10}d(iii), particularly at the end of the domain, which results from spectral leakage in the analytic signal caused by the non-periodicity of the \ac{POD} modes.

\subsection{Propagating Structures in the Wake of a Sphere}

Table~\ref{tab:pairTKE} compares the contributions of the identified \ac{POD} mode pairs and their corresponding \ac{HPOD} modes to the total turbulent kinetic energy (\ac{TKE}).
The first pair of \ac{POD} modes (modes 1 and 2) represents the largest contribution to \ac{TKE}, at 8.24\%, which slightly exceeds the 7.70\% contribution of the first \ac{HPOD} mode.
Subsequent pairs contribute progressively less energy, with the fourth pair (modes 9 and 10) accounting for 2.60\% of the total \ac{TKE}.
The corresponding \ac{HPOD} modes exhibit similar trends, although their contributions differ from those of the \ac{POD} mode pairs.
The first two \ac{HPOD} modes contribute more proportionally to the analytic signal than the corresponding \ac{POD} mode pairs contribute to the original data, whereas the third and fourth \ac{HPOD} modes contribute less than their \ac{POD} counterparts.
This is consistent with the higher-order \ac{POD} modes containing a greater proportion of non-propagating structures, which are excluded from the leading \ac{HPOD} modes.

\begin{table}
  \centering
  \begin{tabular}{c c c c}
    \toprule
    \bf{\ac{HPOD} Mode} & \bf{TKE [\%]} & \bf{\ac{POD} Modes} & \bf{TKE [\%]} \\
    \midrule
    1 & 7.70 & 1, 2 & 8.24 \\
    2 & 4.53 & 4, 5 & 4.75 \\
    3 & 3.70 & 6, 8 & 3.20 \\
    4 & 2.70 & 9,10 & 2.60 \\
    \bottomrule
  \end{tabular}
  \caption{Turbulent kinetic energy contributions of the \ac{HPOD} modes to $\bs{X}^a$ and of the \ac{POD} mode pairs to $\bs{X}$.}\label{tab:pairTKE}
\end{table}

In order to examine the nature and effects of the propagating structures identified from the \ac{HPOD} and \ac{POD} modes, the \ac{HPOD} modes and corresponding \ac{POD} mode pairs were employed to phase-average the instantaneous velocity fluctuations.
The phase-averaged velocity fluctuations based on the phase angle of the $i^{th}$ and $j^{th}$ \ac{POD} modes are defined as
\begin{equation}\label{eq:PhsAvgVelFluc}
  \begin{bmatrix}
    \wt{u'}(\bs{x},\phi) \\ \wt{v'}(\bs{x},\phi)\\
  \end{bmatrix}
  = \frac{1}{N_\phi}\sum_{k=1}^K a_k(t) \psi_k(\bs{x}) \, | \, \phi_{i,j}=\phi,
\end{equation}
where $i$ and $j$ are paired \ac{POD} modes, and $N_\phi$ is the number of instantaneous velocity fields used in the phase average at corresponding phase angle $\phi$.
The mean velocity is combined with the phase-averaged velocity fluctuations to obtain the phase-averaged velocity
\begin{equation}\label{eq:PhsAvgVel}
  \begin{bmatrix}
    \wt{u}(\bs{x},\phi) \\ \wt{v}(\bs{x},\phi)\\
  \end{bmatrix}
  =
  \begin{bmatrix}
    \wt{u'}(\bs{x},\phi) \\ \wt{v'}(\bs{x},\phi)\\
  \end{bmatrix}
  +
  \begin{bmatrix}
    \ol{u}(\bs{x}) \\ \ol{v}(\bs{x})\\
  \end{bmatrix},
\end{equation}
from which the phase-averaged out-of-plane vorticity is computed as
\begin{equation}\label{eq:PhsAvgVrt}
  \wt{\omega}(\bs{x},\phi) = \frac{\partial \wt{v}(\bs{x},\phi)}{\partial x} - \frac{\partial \wt{u}(\bs{x},\phi)}{\partial y}.
\end{equation}
The phase-averaged planar Reynolds stress is obtained by averaging the product of the streamwise and transverse components of the instantaneous velocity fluctuations at a given phase angle
\begin{equation}\label{eq:PhsAvgReStress}
  \wt{u'v'}(\bs{x},\phi) = \frac{1}{N_\phi}\sum_{k=1}^K \big( a_k(t) \psi^u_k(\bs{x}) \big) \big( a_k(t) \psi^v_k(\bs{x}) \big) | \, \phi_{i,j}=\phi,
\end{equation}
where $\psi^u_k(\bs{x})$ and $\psi^v_k(\bs{x})$ are the streamwise and transverse components of the $k^{th}$ \ac{POD} mode, respectively.
The \ac{TKE} of the phase-averaged velocity is given by
\begin{equation}\label{eq:PhsAvgTKE}
  \wt{\operatorname{TKE}}(\bs{x},\phi) = \big( \wt{u'}(\bs{x},\phi) \big)^2 + \big( \wt{v'}(\bs{x},\phi) \big)^2.
\end{equation}

\begin{figure}
  \centering
  \includegraphics[width=5.3in]{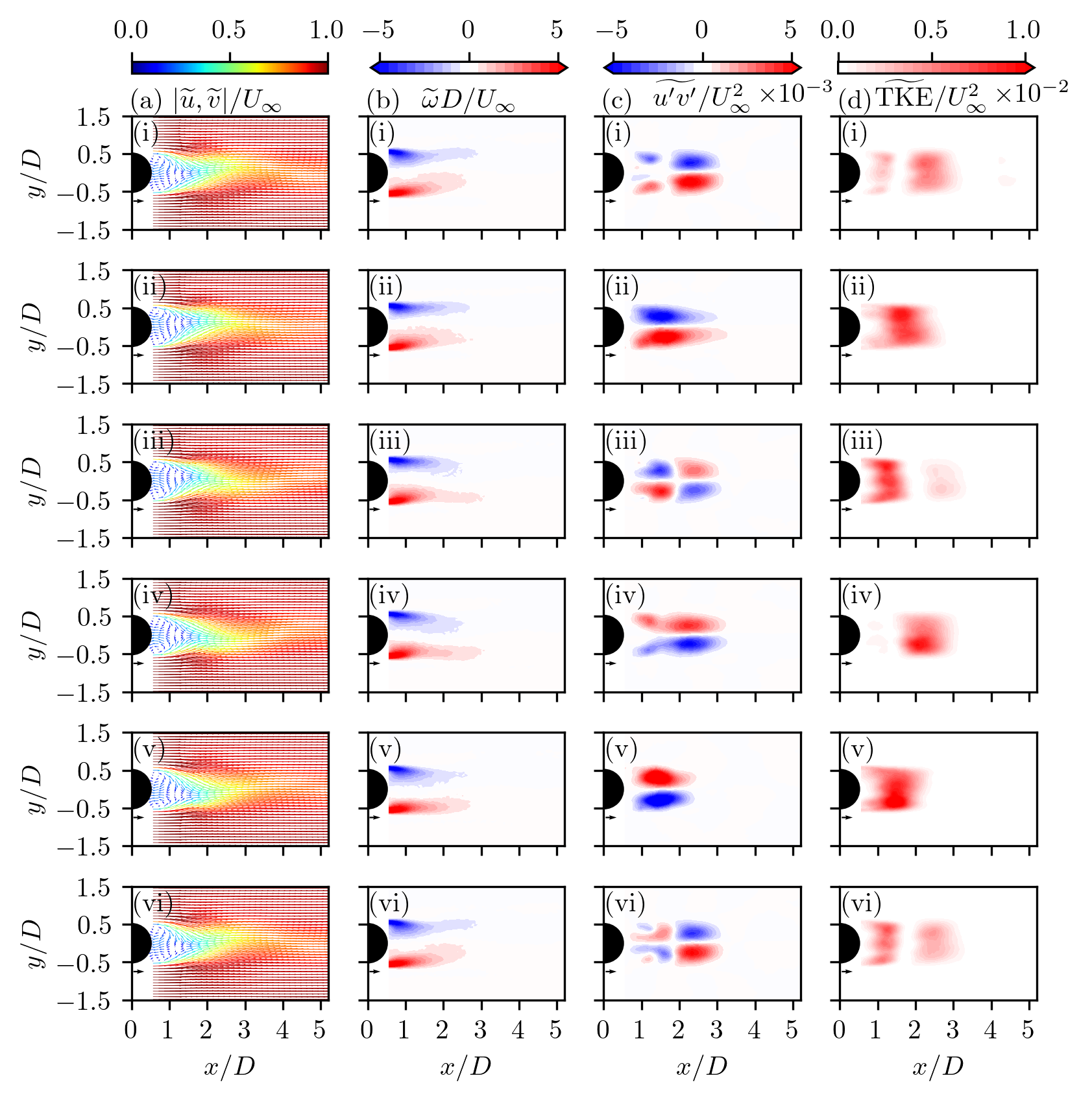}
  \caption{Phase-averaged (a) velocity, (b) vorticity, (c) planar Reynolds stress, and (d) \ac{TKE} using the first and second \ac{POD} modes at $\phi=$ (i) 0, (ii) $\pi/6$, (iii) $\pi/3$, (iv) $\pi/2$, (v) $2\pi/3$, and (vi) $5\pi/6$.
  See supplementary movie 9 for an animated version of this figure.}\label{fig:POD1,2}
\end{figure}

\begin{figure}
  \centering
  \includegraphics[width=5.3in]{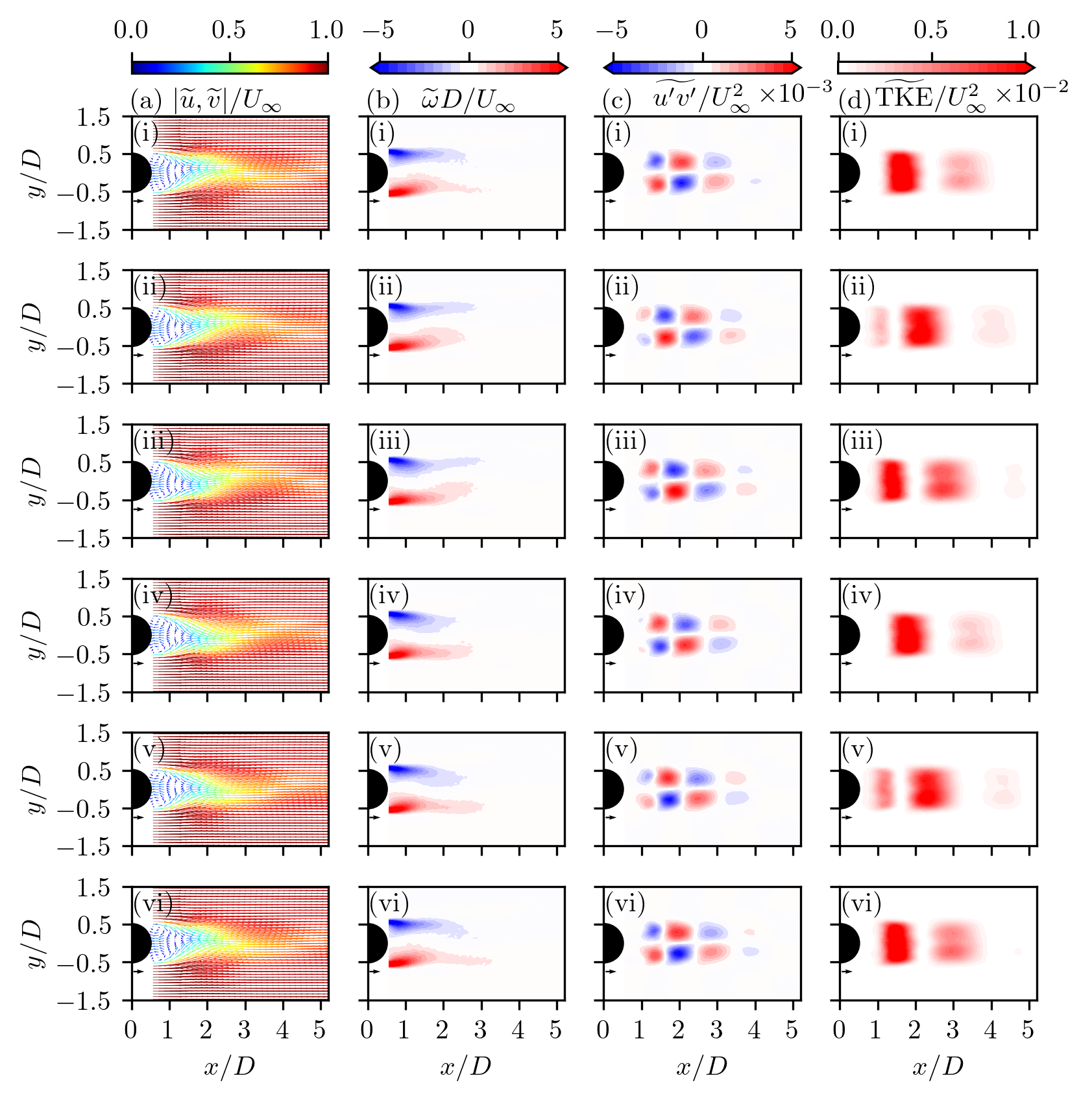}
  \caption{(a) Velocity, (b) vorticity, (c) planar Reynolds stress, and (d) \ac{TKE} of the first \ac{HPOD} mode at $\phi=$ (i) 0, (ii) $\pi/6$, (iii) $\pi/3$, (iv) $\pi/2$, (v) $2\pi/3$, and (vi) $5\pi/6$.
  See supplementary movie 10 for an animated version of this figure.}\label{fig:HPOD1}
\end{figure}

The first and second \ac{POD} modes, shown in figure~\ref{fig:POD1,2}, and the first \ac{HPOD} mode, shown in figure~\ref{fig:HPOD1}, represent a flapping motion in the wake.
The phase-averaged velocity obtained using the first and second \ac{POD} modes and the first \ac{HPOD} mode, shown in figure~\ref{fig:POD1,2}a and~\ref{fig:HPOD1}a, respectively, are qualitatively indistinguishable, as is the phase-averaged vorticity, shown in figures~\ref{fig:POD1,2}b and~\ref{fig:HPOD1}b, respectively.
The structures in the planar Reynolds stresses from the phase-averaged \ac{POD} modes, shown in figure~\ref{fig:POD1,2}c, are longer and more intense for $x/D < 3$ than those in the first \ac{HPOD} mode, shown in figure~\ref{fig:HPOD1}c, and disappear after $x/D = 3$.
The structures in the planar Reynolds stress associated with the \ac{HPOD} mode are shorter and more consistent in size across the measured domain.
The intensity of the structures is lower than that of the \ac{POD} modes, but decays more slowly, with structure remaining visible beyond $x/D = 3$.
The phase-averaged \ac{TKE} of the \ac{POD} modes, shown in figure~\ref{fig:POD1,2}d, appears in periodically shed packets that increase in intensity until $x/D \approx 3$, and then decay after $x/D \approx 3$.
The \ac{TKE} of the \ac{HPOD} mode, shown in figure~\ref{fig:HPOD1}d, exhibits the same periodic structures as the \ac{POD} modes but with greater intensity.
This is consistent with the \ac{TKE} of the analytic signal of the turbulent fluctuations being significantly greater than the actual turbulent fluctuations.

\begin{figure}
  \centering
  \includegraphics[width=5.3in]{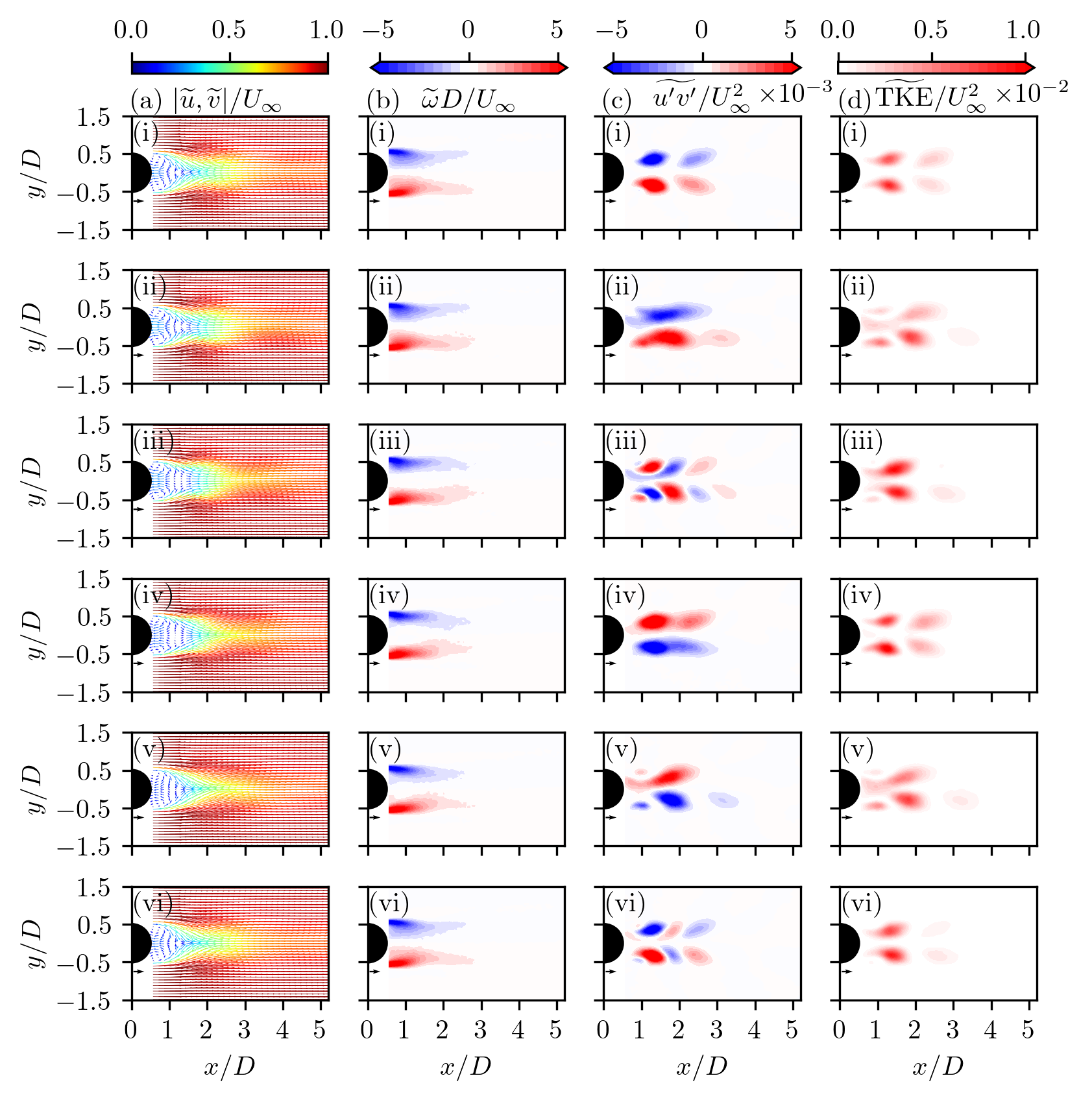}
  \caption{Phase-averaged (a) velocity, (b) vorticity, (c) planar Reynolds stress, and (d) \ac{TKE} using the fourth and fifth \ac{POD} modes at $\phi=$ (i) 0, (ii) $\pi/6$, (iii) $\pi/3$, (iv) $\pi/2$, (v) $2\pi/3$, and (vi) $5\pi/6$.
  See supplementary movie 11 for an animated version of this figure.}\label{fig:POD4,5}
\end{figure}

\begin{figure}
  \centering
  \includegraphics[width=5.3in]{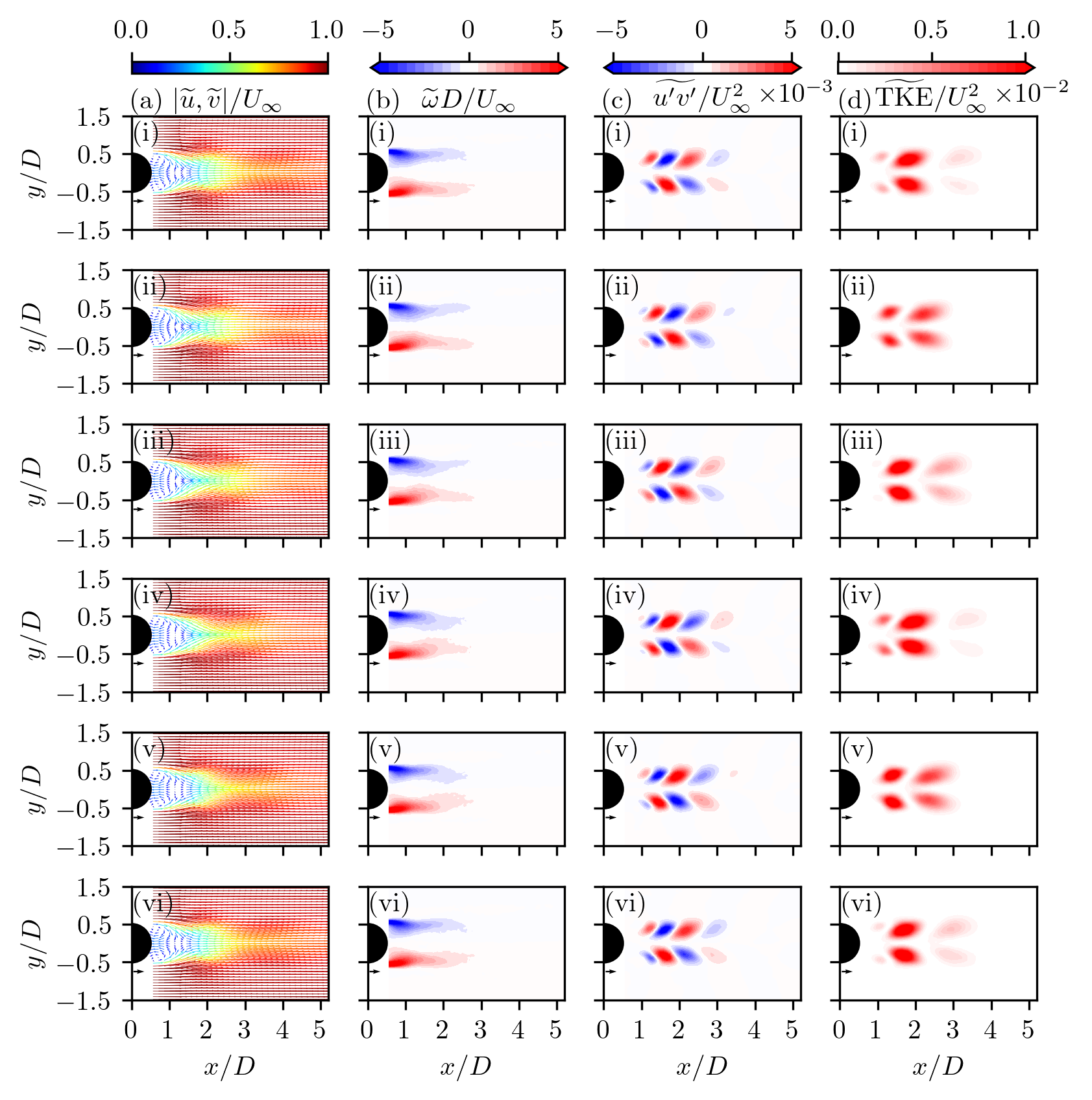}
  \caption{(a) Velocity, (b) vorticity, (c) planar Reynolds stress, and (d) \ac{TKE} of the second \ac{HPOD} mode at $\phi=$ (i) 0, (ii) $\pi/6$, (iii) $\pi/3$, (iv) $\pi/2$, (v) $2\pi/3$, and (vi) $5\pi/6$.
  See supplementary movie 12 for an animated version of this figure.}\label{fig:HPOD2}
\end{figure}

The structure represented by the fourth and fifth \ac{POD} modes and the second \ac{HPOD} mode corresponds to a pulsation in the streamwise fluctuations, as evidenced by the deformation of the low-velocity region behind the sphere, shown in figure~\ref{fig:POD4,5}a and~\ref{fig:HPOD2}a, respectively.
The streamwise pulsation is also evident in the phase-averaged out-of-plane vorticity, where the high-vorticity regions on either side of the sphere extend further toward the centreline near the sphere, as shown in figures~\ref{fig:POD4,5}b(ii) and~\ref{fig:HPOD2}b(ii).
This extension toward the centreline occurs at approximately $x/D = 2$, as shown in figures~\ref{fig:POD4,5}b(v) and~\ref{fig:HPOD2}b(v).
While the structures in the planar Reynolds stress vary considerably in shape and size in the phase-averaged \ac{POD} modes, shown in figure~\ref{fig:POD4,5}c, they remain relatively uniform in size and shape in the \ac{HPOD} mode, as shown in figure~\ref{fig:HPOD2}c.
The size and shape of the phase-averaged \ac{TKE} associated with the \ac{POD} modes, shown in figure~\ref{fig:POD4,5}d, are more uniform than those of the planar Reynolds stresses.
These structures are similar to, but less planar-symmetric than, those in the phase-averaged \ac{TKE} of the \ac{HPOD} mode, which also exhibit greater intensity and a more uniform pattern.
The intensity of the \ac{TKE} increases immediately behind the sphere up to $x/D = 2$, as seen in figure~\ref{fig:HPOD2}d(ii), before decreasing further downstream, as seen in figure~\ref{fig:HPOD2}d(iv).

\begin{figure}
  \centering
  \includegraphics[width=5.3in]{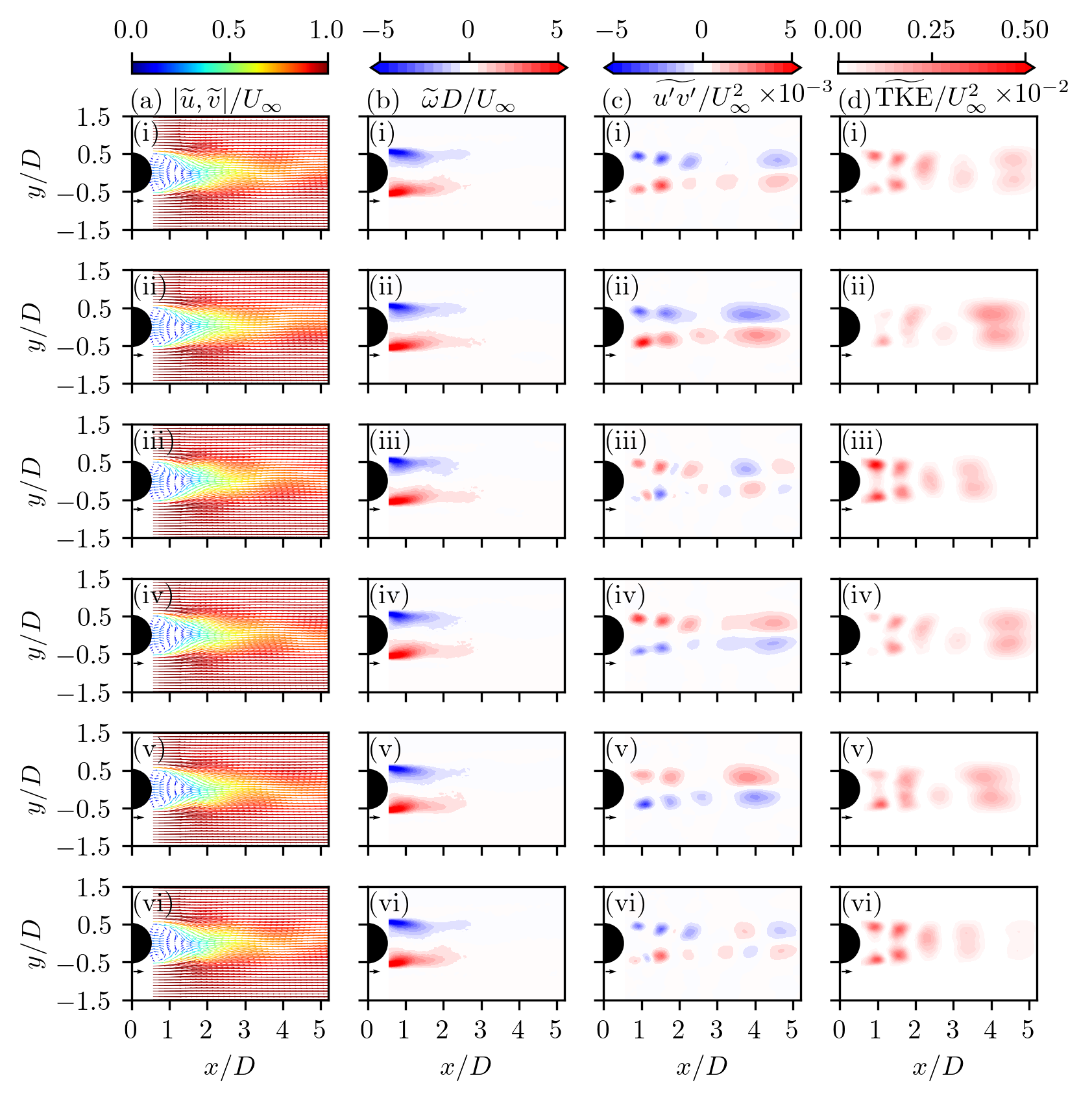}
  \caption{Phase-averaged (a) velocity, (b) vorticity, (c) planar Reynolds stress, and (d) \ac{TKE} using the sixth and eighth \ac{POD} modes at $\phi=$ (i) 0, (ii) $\pi/6$, (iii) $\pi/3$, (iv) $\pi/2$, (v) $2\pi/3$, and (vi) $5\pi/6$.
  See supplementary movie 13 for an animated version of this figure.}\label{fig:POD6,8}
\end{figure}

\begin{figure}
  \centering
  \includegraphics[width=5.3in]{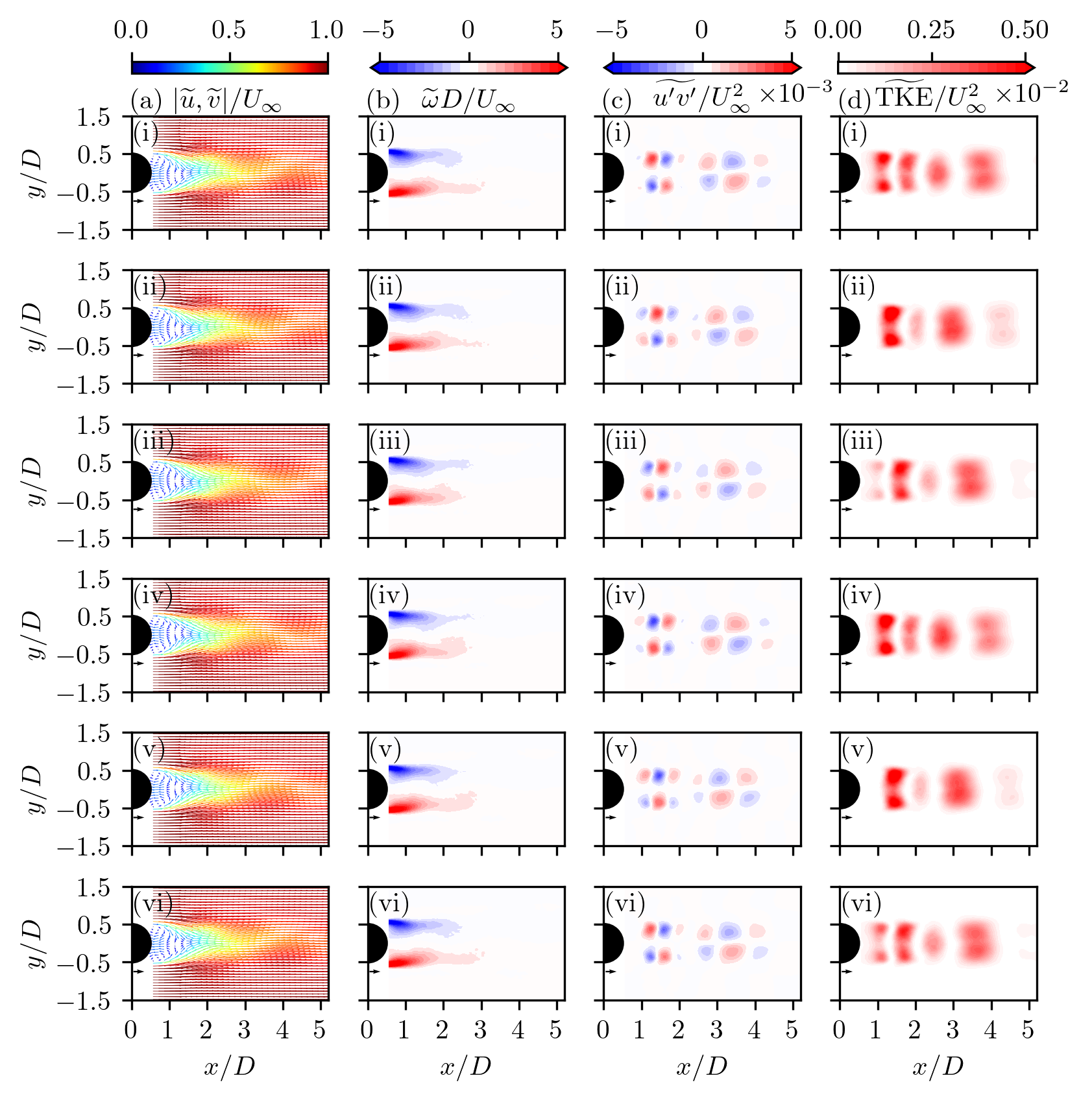}
  \caption{(a) Velocity, (b) vorticity, (c) planar Reynolds stress, and (d) \ac{TKE} of the third \ac{HPOD} mode at $\phi=$ (i) 0, (ii) $\pi/6$, (iii) $\pi/3$, (iv) $\pi/2$, (v) $2\pi/3$, and (vi) $5\pi/6$.
  See supplementary movie 14 for an animated version of this figure.}\label{fig:HPOD3}
\end{figure}

The sixth and eighth \ac{POD} modes correspond to a flapping motion in the wake, similar to that of the first and second \ac{POD} modes, but with a shorter wavelength.
Consequently, the phase-averaged velocity, shown in figure~\ref{fig:POD6,8}a, and phase-averaged vorticity, shown in figure~\ref{fig:POD6,8}b, closely resemble those of the first and second \ac{POD} modes.
The smaller structures are evident in the phase-averaged planar Reynolds stress, shown in figure~\ref{fig:POD6,8}c, particularly near the sphere.
Further downstream of the sphere, the structures increase in streamwise extent, as illustrated in figure~\ref{fig:POD6,8}c(iv).
The smaller structures in the phase-averaged \ac{TKE}, shown in figure~\ref{fig:POD6,8}d, are more distinctly separated across the sphere's centreline than in the first and second \ac{POD} modes, where larger structures extend to the centreline.
The structures in the phase-averaged planar Reynolds stresses of the third \ac{HPOD} mode, shown in figure~\ref{fig:HPOD3}c, exhibit more uniformity in size, shape and arrangement compared to those of the \ac{POD} modes.
Additionally, a region with low planar Reynolds stress occurs around $x/D = 2$ for all phase angles shown in figure~\ref{fig:HPOD3}b, corresponding to smaller structures in the streamwise turbulent fluctuations in this region, shown in figure~\ref{fig:HPOD3_POD6,8}b.
The structures in the phase-averaged \ac{TKE} are more distinctly defined for the \ac{HPOD} modes, as shown in figure~\ref{fig:HPOD3}d.

\begin{figure}
  \centering
  \includegraphics[width=5.3in]{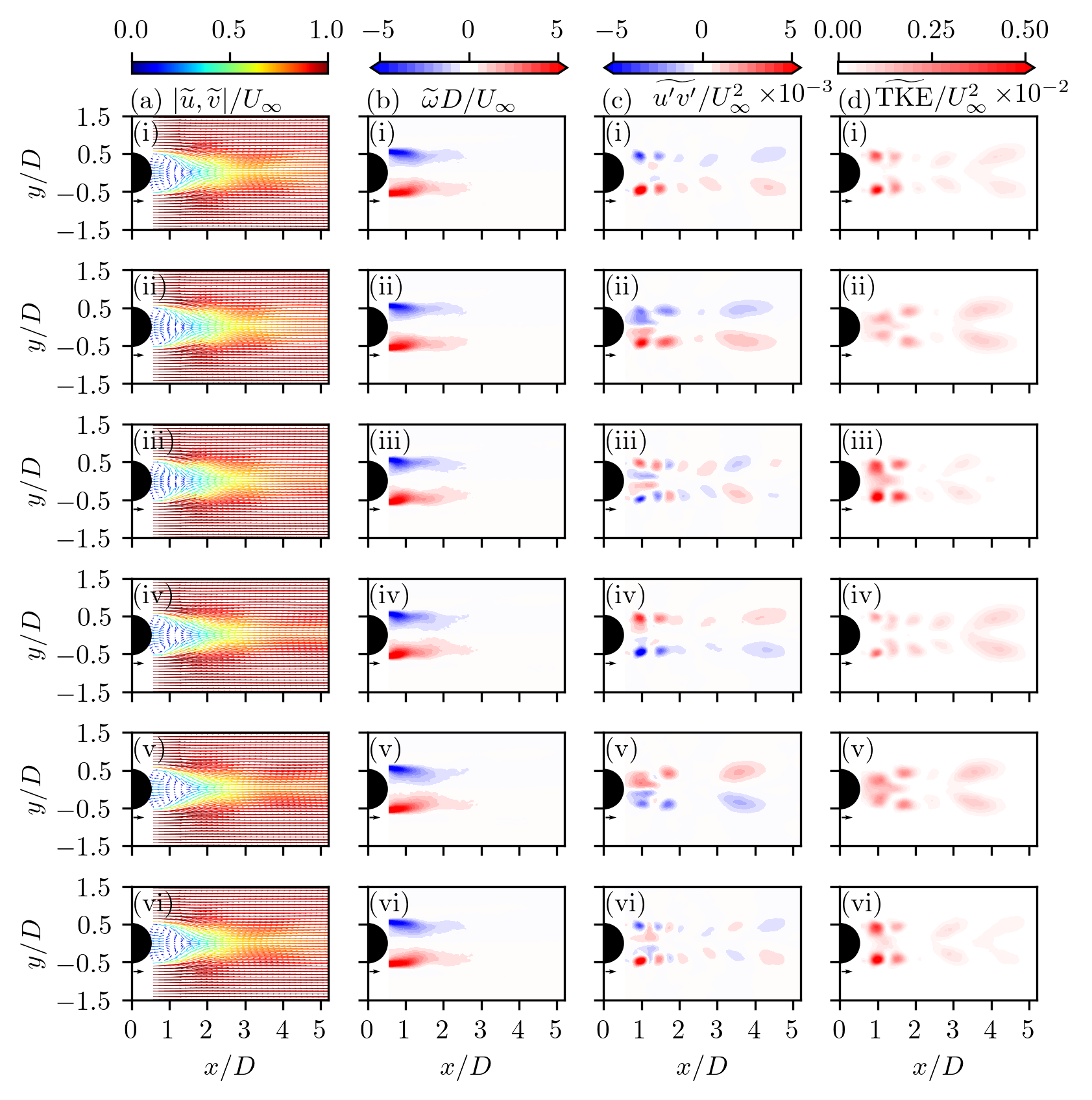}
  \caption{Phase-averaged (a) velocity, (b) vorticity, (c) planar Reynolds stress, and (d) \ac{TKE} using the ninth and tenth \ac{POD} modes at $\phi=$ (i) 0, (ii) $\pi/6$, (iii) $\pi/3$, (iv) $\pi/2$, (v) $2\pi/3$, and (vi) $5\pi/6$.
  See supplementary movie 15 for an animated version of this figure.}\label{fig:POD9,10}
\end{figure}

\begin{figure}
  \centering
  \includegraphics[width=5.3in]{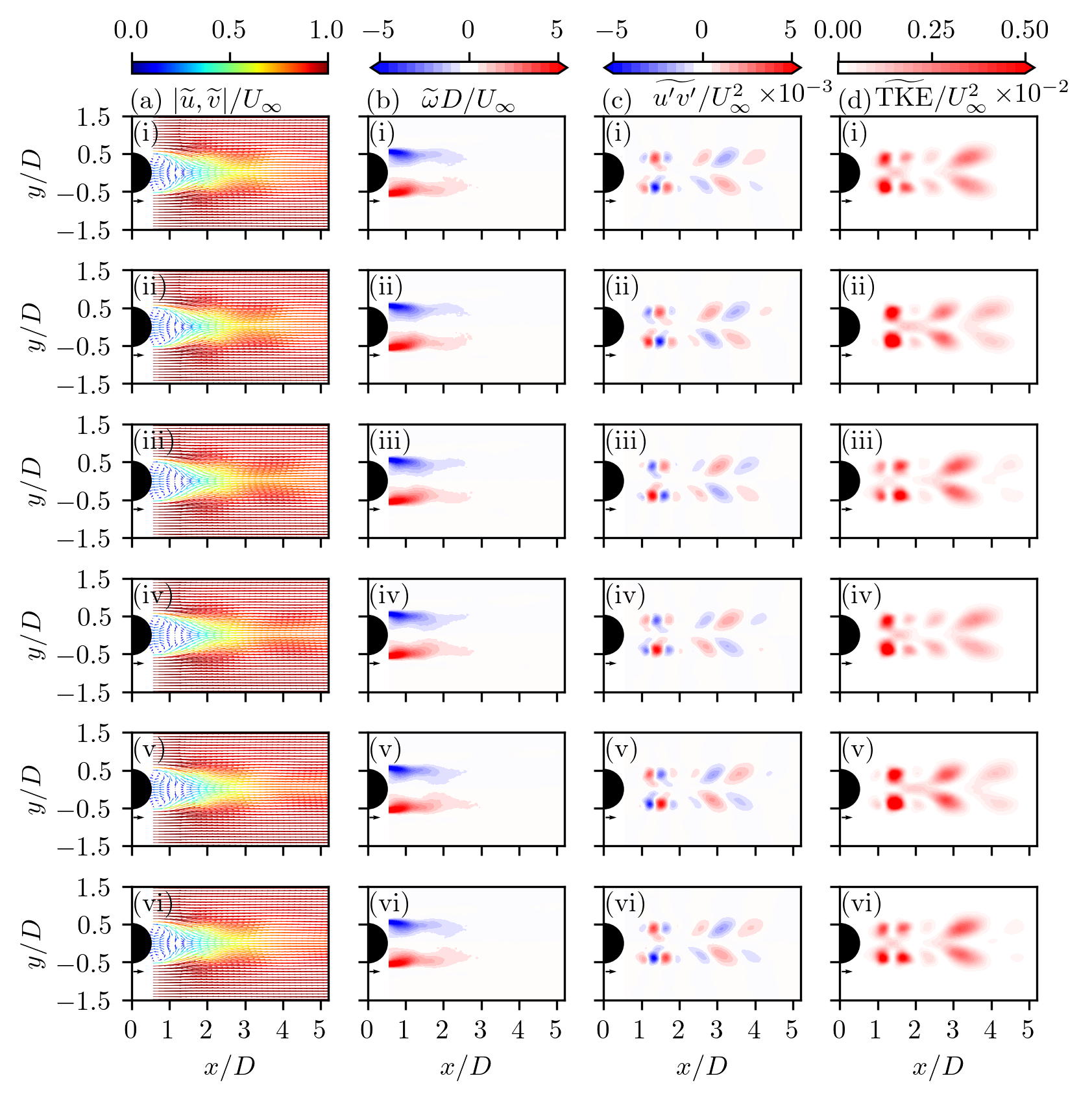}
  \caption{(a) Velocity, (b) vorticity, (c) planar Reynolds stress, and (d) \ac{TKE} of the fourth \ac{HPOD} mode at $\phi=$ (i) 0, (ii) $\pi/6$, (iii) $\pi/3$, (iv) $\pi/2$, (v) $2\pi/3$, and (vi) $5\pi/6$.
  See supplementary movie 16 for an animated version of this figure.}\label{fig:HPOD4}
\end{figure}

The structure represented by the ninth and tenth \ac{POD} modes and the fourth \ac{HPOD} mode resembles that of the fourth and fifth \ac{POD} modes and the second \ac{HPOD} mode, namely a periodic streamwise pulsation in the wake.
The phase-averaged velocity and vorticity for the \ac{POD} modes and \ac{HPOD} mode, shown in figures~\ref{fig:POD9,10}a--b and~\ref{fig:HPOD4}a--b, respectively, closely resemble those of the lower-order modes.
The structures in the planar Reynolds stress phase-averaged using the \ac{POD} modes, shown in figure~\ref{fig:POD9,10}c, are small and rounded in the near wake of the sphere but become larger and more ovular further downstream, with the latter angled away from the centreline in alignment with the V-shaped structures in the \ac{POD} modes.
While the arrangement of these structures varies considerably near the sphere in the \ac{POD} mode phase average, the planar Reynolds stress of the \ac{HPOD} mode, shown in figure~\ref{fig:HPOD4}c, clearly exhibits rounded structures for $x/D < 2$ and more V-shaped structures for $x/D > 2$. 
Both the \ac{POD} modes and \ac{HPOD} modes exhibit a pronounced reduction in the planar Reynolds stress near $x/D = 2$, which separates the rounded structures from the V-shaped ones.
However, the \ac{HPOD} modes display this transition more prominently, whereas the \ac{POD} modes exhibit additional structures in the very near wake which are absent from the \ac{HPOD} modes.
The transition from round structures near the sphere to V-shaped structures further downstream is evident in the phase-averaged \ac{TKE}, shown in figure~\ref{fig:POD9,10}d and~\ref{fig:HPOD4}d, for the \ac{POD} modes and the \ac{HPOD} mode, respectively.
This also indicates that the structures near the sphere are more intense and decay towards $x/D = 2$, whereas the V-shaped structures increase in intensity around $x/D = 3$ before gradually weakening downstream.
Similar to the planar Reynolds stress, the structures in the \ac{TKE} are more distinctly defined in the \ac{HPOD} mode due to the reduced significance of non-propagating structures in the analytic signal.

The structural details revealed by the \ac{POD} mode pairs identified from the \ac{HPOD} modes differ from those of the corresponding \ac{HPOD} modes.
The primary distinction between these methods stems from emphasis on propagating structures in the leading \ac{HPOD} modes, whereas non-propagating structures are relegated to higher-order modes.
Phase-averaging the resulting modes yields more uniformly shaped structures, thereby facilitating the identification and interpretation of propagating structures.
However, the reduced emphasis on non-propagating structures may result in a less accurate representation of flow features that do not propagate in the direction of the Hilbert transform, such as those confined to the near wake of the sphere.
Although applying the Hilbert transform directly to the \ac{POD} modes is computationally more efficient than performing \ac{HPOD} to pair \ac{POD} modes, it also artificially imposes propagation on structures that may not physically propagate.
Thus, both methods of pairing \ac{POD} modes introduce non-physical assumptions into the flow analysis, which should be carefully considered when interpreting the results.

\section{Concluding Remarks}\label{sec:Conclusions}

The velocity in the wake of a sphere at $Re_D = 7780$ was measured using \ac{2C-2D} \ac{MCCD}\ac{PIV}, with \ac{POD} and \ac{HPOD} applied to the turbulent fluctuations.
The resulting \ac{POD} and \ac{HPOD} modes were compared to assess whether the \ac{HPOD} modes, which emphasise propagating structures, could be used to identify \ac{POD} mode pairs representing the same structure at different phase angles.
The analytic signal of the \ac{POD} modes was then employed as a substitute for the \ac{HPOD} modes to identify paired \ac{POD} modes without performing the computationally intensive \ac{HPOD}.
The \ac{POD} mode pairs identified using these methods were then used to phase-average the turbulent fluctuations to enable the examination of propagating structures in the sphere's wake.

The leading \ac{HPOD} modes are shown to represent the same structures as the leading \ac{POD} modes, with a greater emphasis on the propagating structures in the former.
While each \ac{HPOD} mode consists of a real and imaginary part corresponding to a phase-shift of $\pi/2$ in the mode, this comes at the cost of spectral leakage due to the assumption of periodicity imposed by the Hilbert transform.
As a result, it can be difficult to differentiate between genuine flow structures and artefacts introduced by the transform, particularly when taking two-dimensional measurements of a three-dimensional flow.
Performing the Hilbert transform on the \ac{POD} modes to create an analytic signal of the \ac{POD} modes is shown to identify the same \ac{POD} mode pairs as the \ac{HPOD} modes.
This method allows the same structures to be identified, without introducing the non-physical elements introduced by the \ac{HPOD}.
The identified \ac{POD} mode pairs represent flapping and pulsating motions in the wake of the sphere.

As the velocity fields in the present study are not temporally resolved, the Hilbert transform was applied in the streamwise direction, meaning the identified structures are those propagating along this axis.
As a result, the temporal information of the structures is not captured by either decomposition.
However, the paired \ac{POD} modes offer insight into the dynamics of turbulent structures, even when flow complexity makes it difficult to identify corresponding \ac{POD} modes.

\section*{Acknowledgements}
  Shaun Davey gratefully acknowledges the support of the Australian Commonwealth Government through a Research Training Program (RTP) Scholarship.
  This research was undertaken using resources provided through a Monash HPC-NCI Merit Allocation from the National Computational Infrastructure (NCI Australia), an NCRIS-enabled capability supported by the Australian Government.
  The authors also acknowledge the support of the Australian Research Council (ARC) in funding this research through Linkage Infrastructure, Equipment and Facilities (LIEF) grants LE100100222 and LE180100166.

\section*{Declaration of Interests}
  The authors report no conflict of interest.

\bibliographystyle{ieeetr}
\bibliography{refs}

\end{document}